\documentclass[a4paper,fleqn]{cas-sc}

\usepackage[authoryear,longnamesfirst]{natbib}
\usepackage[section]{placeins}

\def\tsc#1{\csdef{#1}{\textsc{\lowercase{#1}}\xspace}}
\tsc{WGM}
\tsc{QE}

\ExplSyntaxOn
\cs_set:Npn \__first_footerline: {\group_begin:\small\sffamily Qu~et~al.:~Preprint\group_end:}
\ExplSyntaxOff
\AtBeginDocument{\hypersetup{
  pdfauthor={Hong Qu; Zhaoxiang Xu; Jinbo Luo; Yujie Zhao; Jie Zhang; Yadie Yang},
  pdftitle={EasyFashion: A Human-AI Co-Creation System for Personalized Fashion Design and Sewing Pattern Generation},
  pdfsubject={Human-AI co-creation for personalized fashion design},
  pdfkeywords={Human-AI Co-Creation, Personalized Fashion Design, Virtual Try-On, Sewing Pattern Generation}
}}

\begin{document}
\let\WriteBookmarks\relax
\def\floatpagepagefraction{1}
\def\textpagefraction{.001}

\shorttitle{EasyFashion}

\shortauthors{Qu et al.}

\title [mode = title]{EasyFashion: A Human-AI Co-Creation System for Personalized Fashion Design and Sewing Pattern Generation}

\author[1]{Hong Qu}[orcid=0000-0003-3484-0019]
\author[1]{Zhaoxiang Xu}[orcid=0009-0003-4235-9257]
\author[1]{Jinbo Luo}[orcid=0009-0005-8793-4520]
\author[2]{Yujie Zhao}[orcid=0009-0007-2963-8306]
\author[3]{Jie Zhang}[orcid=0000-0001-8219-5590]
\cormark[1]
\author[1]{Yadie Yang}[orcid=0000-0002-1985-7552]
\cormark[1]
\affiliation[1]{
    organization={The School of Fashion and Textiles, The Hong Kong Polytechnic University},
    city={Hong Kong SAR},
    country={China}
}

\affiliation[2]{
    organization={The School of Arts and Design, University of Sanya},
    country={China}
}

\affiliation[3]{
    organization={Faculty of Applied Sciences, Macao Polytechnic University},
    city={Macao SAR},
    country={China}
}

\cortext[1]{Co-corresponding authors.}

\begin{abstract}
People often want garments that reflect their aesthetic preferences, fit their bodies, and meet their sizing needs, yet turning these requirements into physical garments remains difficult. Ready-to-wear options provide limited personalization, while custom tailoring is costly and time-consuming. Recent generative artificial intelligence (AI) systems can visualize garment ideas but often stop short of supporting downstream production. To address this gap, we present EasyFashion, a human-AI co-creation system that enables users to iteratively refine design intent for personalized garment style and size, evaluate designs through virtual try-on on reconstructed personal avatars, and generate sewing patterns for garment production. Using reference images, text descriptions, and body photos as input, EasyFashion translates user intent into structured garment specifications and try-on results. Technical experiments, user studies, and a real-world production case demonstrate the value of EasyFashion for multimodal design expression, body-specific evaluation, and production-oriented outputs in personalized garment design. Our project page is available at \url{https://zoequ.github.io/easyfashion.github.io/}.
\end{abstract}

\begin{keywords}
 Human-AI Co-Creation\sep Personalized Fashion Design\sep Multimodal Interaction \sep Virtual Try-On \sep Human Body Reconstruction\sep Sewing Pattern Generation\sep Vision-Language Models
\end{keywords}

\maketitle
\hypersetup{pdfauthor={Hong Qu; Zhaoxiang Xu; Jinbo Luo; Yujie Zhao; Jie Zhang; Yadie Yang},pdfsubject={Human-AI co-creation for personalized fashion design}}

\section{Introduction}

Choosing and purchasing garments is a routine part of everyday life, yet the process is often frustrating. Consumers care deeply about how a garment will look and fit on their own body, but they usually have limited information before making a decision. This mismatch frequently leads to dissatisfaction, particularly with respect to fit and style expectations, and is a major driver of product returns in online fashion retail~\citep{JiayinLi2024, wang_sf_2023, Diggins2016, stocker2021, gry2023advances}.
Personalized garment design is therefore appealing because it offers the possibility of better aligning garments with individual body characteristics and personal preferences. However, for non-professional users, pursuing such personalization in practice remains difficult.

This challenge is important because personalized garment addresses a genuine everyday need. People often want garments that better reflect their aesthetic preferences while fitting their own body more appropriately~\citep{masscustomization2002,JiayinLi2024,wang_sf_2023}. In general, these needs involve two complementary dimensions: appearance personalization, such as style, color, and pattern, and fit personalization, such as garment length, overall looseness or tightness, and fit adjustments in specific body areas. Despite this demand, existing customization workflows still rely heavily on professional tailors or designers, and the process is often time-consuming, costly, and difficult to revise once production has begun~\citep{claudia2000,eckert2001,gill2015}.

A central obstacle in this process is the communication of design intent. For non-professional users, the primary difficulty is often not garment production itself, but articulating what they want in a form that can be accurately interpreted and translated into a realizable garment. In practice, users tend to rely on everyday and often ambiguous expressions, such as ``not too tight,'' ``a bit longer,'' ``make me look slimmer,'' or ``like this photo, but less revealing.'' They may also be uncertain about which styles suit them, while custom orders are often perceived as risky because garments are difficult to revise, undo, or return once made. As a result, unclear intent frequently leads to repeated back-and-forth communication, multiple revisions, and production rework, which increase both time and cost~\citep{eckert2001, claudia2000, masscustomization2002}.
These barriers discourage exploration and limit the accessibility of personalized garment design.

\begin{figure}[pos=h]
  \centering
  \includegraphics[width=\linewidth]{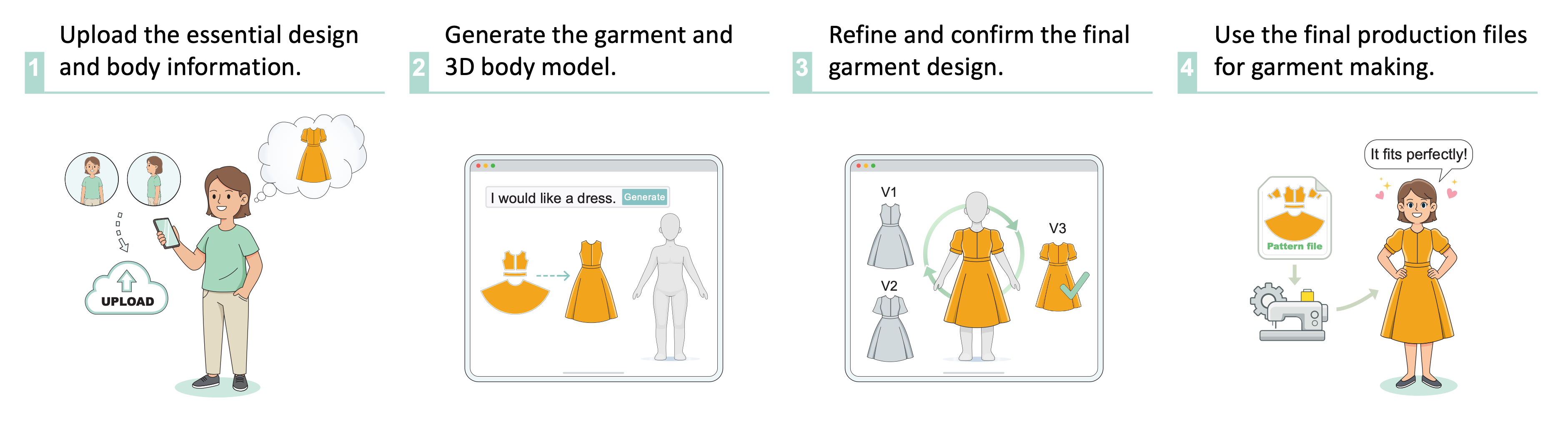}
  \caption{Illustration of the user journey in our EasyFashion system. The workflow consists of four stages: (1) Upload the essential design and body information (e.g., reference images and body photos); (2) Generate the garment and 3D body model through the system; (3) Refine and confirm the final garment design; and (4) Use the final production files for garment making.}
  \label{fig:storying}
  
\end{figure}

Recent advances in AI, particularly vision-language models (VLMs) and generative fashion pipelines, have made it increasingly feasible for users to express design intent through text, reference images, and sketches, and to receive outputs such as generated fashion images or virtual try-on previews~\citep{fabricdiff2024, xu2025ootdiffusion, gaussiangarment2025, dresscode2024, chatgarment2025}.
These developments create an opportunity to make personalized garment design more accessible by helping users externalize their preferences and preview possible outcomes before committing to production decisions~\citep{fashionQ2021,designprompt2024,sketchai2025}. However, in practice, many existing AI-based fashion systems still emphasize visual generation and style variation, such as designer-inspired image synthesis, texture transfer, and virtual try-on~\citep{styleme2023,fabricdiff2024,xu2025ootdiffusion,virtual_tryon_review2024}, rather than supporting users in making confident judgments about how a garment would actually look and fit on their own body~\citep{styleme2023, fabricdiff2024, xu2025ootdiffusion}.

Moreover, current creative workflows still present substantial friction for non-professional users. First, many systems depend on long and highly detailed text prompts, which can be burdensome for users who do not know the appropriate terminology~\citep{ai_or_me_chi2024, designprompt2024}.
Second, when system feedback is slow or difficult to interpret, users may lose momentum during exploration and become less willing to continue refining their ideas~\citep{impact_ai2024, keep_users_in_the_flow2015, time_constraints2016}.
Third, even when a system generates visually appealing garment images, those images alone are often insufficient for judging fit on one's own body, comparing subtle design changes, or converging toward production-oriented garment specifications. Meanwhile, virtual try-on results are not inherently reliable, since errors in body reconstruction, garment simulation, and rendering may mislead users~\citep{virtual_tryon_review2024,Kahr2024}.
Taken together, these limitations point to an important Human-Computer Interaction (HCI) opportunity: to support low-burden intent expression and meaningful iterative refinement through multimodal human-AI co-creation. In this paper, we define multimodal human-AI co-creation as an iterative workflow in which users express and refine garment design intent through text, reference images, and optional sketches, while the system maintains an editable garment specification that supports both body-specific virtual try-on and production-oriented sewing pattern generation.

To better understand these needs, we conducted a formative study involving both non-professional users and professionals through a questionnaire survey and semi-structured interviews. Our findings suggest that interactive tools for personalized garment design should support low-burden input that aligns with how people naturally describe garments, enable iterative refinement with feedback that is sufficiently responsive for continued exploration, and provide body-specific evaluation, such as virtual try-on, to help users assess fit and appearance on themselves rather than on an abstract avatar. These insights motivate our overall goal: to make personalized garment design more accessible to non-professional users by enabling multimodal intent input, supporting iterative refinement, and linking conceptual design intent to production-oriented outputs through body-specific feedback.

Based on these insights, we present \textbf{EasyFashion} (Fig.~\ref{fig:storying}), an end-to-end interactive system for personalized garment design. At the technical level, EasyFashion is built around two key components. The first is a VLM-based Fashion Agent, which translates multimodal user inputs, including text, design reference images, and sketches, into a structured and editable garment specification in JavaScript Object Notation (JSON) format, and supports iterative refinement by updating this specification in response to user feedback. In addition to these design-intent inputs, users also provide a small set of personal photos used exclusively for body reconstruction. The second component is a human body reconstruction module built on Skinned Multi-Person Linear Model (SMPL)~\citep{smpl2015}. It generates a personalized body model from these user photos, thereby enabling body-specific virtual try-on. Importantly, these two components are modular: the Fashion Agent does not control body reconstruction, and the reconstructed body serves as an independent, user-specific basis for evaluation. Downstream modules then use the structured garment specification to generate digital fabric appearance, produce sewing patterns, and render 3D try-on results~\citep{neuraltailor2022, garmentcode2023}. Our primary focus is on the interactive experience enabled by the Fashion Agent and on improving the reliability of body-specific try-on through robust body reconstruction. We further examine the feasibility of the proposed workflow through physical prototyping of selected garments.

Accordingly, this paper investigates the following \textbf{research questions (RQs)}:

\begin{itemize}
  
   \item \textbf{RQ1:} \textit{(System Design)}: How can a personalized fashion design system (EasyFashion)  be designed for non-professional users?
  \item \textbf{RQ2:} \textit{(System Validation)}: Is EasyFashion effective for supporting personalized fashion design by non-professional users?
  \item \textbf{RQ3:} \textit{(System Validation)}: Is EasyFashion user-friendly for supporting personalized fashion design by non-professional users?
\end{itemize}

To answer these questions, we evaluate EasyFashion in two stages. First, we conduct experiments to assess key technical components, including: (i) the robustness of human body reconstruction under varying user photo conditions, and (ii) the ability of the Fashion Agent to translate different input conditions into complete and editable garment specifications, as reflected in the constraint satisfaction and output consistency of the generated garments. Second, we conduct a user study to evaluate usability, perceived ease of use, perceived acceptability of the system's responsiveness, and whether the workflow helps users iteratively refine their designs toward production-oriented specifications with the support of body-specific try-on feedback. Standard questionnaires and subjective ratings are used for evaluation~\citep{pssuq1992, sus1996, muralart2025}. 

This paper makes three main contributions:

\begin{itemize}
    \item We propose EasyFashion, an end-to-end interactive system that supports multimodal design intent input and iterative refinement through a VLM-based Fashion Agent, personalized body reconstruction, body-specific virtual try-on feedback, and sewing pattern generation for production.
    
    \item We conduct a formative study that characterizes how non-professional users communicate garment design intent, why intent expression and iterative refinement are challenging, and what support is needed for low-burden input and body-specific evaluation.

  \item We present technical and user evaluations showing how different input conditions affect design intent communication and output quality, and how multimodal human-AI co-creation can help non-professional users iteratively transform conceptual design ideas into production-oriented specifications and sewing patterns, supported by body-specific try-on feedback and validated through physical prototyping.
\end{itemize}

\section{Related Works}
This section reviews prior work that informs the design of EasyFashion from three perspectives: multimodal human-AI co-creation in fashion design, sewing pattern-based garment generation, and human body reconstruction for virtual try-on. Together, these strands establish the technical and interactional foundations of our system.

\subsection{Multimodal Human-AI Co-Creation in Fashion Design}

Human-AI collaboration has broadened access to fashion ideation and customization by supporting tasks such as inspiration retrieval, sketch assistance, digital print generation, and style exploration~\citep{COFI2023, human-ai-co-creation2024, reflexive_data_curation2024, zou2025, guo_review_fashion_design2023, aifashion2024}.
Recent generative models, including DALL-E~\citep{dalle2021}, Stable Diffusion~\citep{stablediffusion2022}, and Imagen~\citep{imagen2022} can generate high-quality fashion images, while domain-specific systems such as Fashion-GPT~\citep{fashiongpt2023} and StyleMe~\citep{styleme2023} further support style exploration and design variation. In this paper, we use multimodal human-AI co-creation to describe workflows in which users communicate design intent through text, reference images, and sketches, and AI systems generate outputs that support the design process. Despite these advances, intent communication remains difficult for non-professional users, because their inputs are often vague, incomplete, or not easily translated into concrete design decisions~\citep{liu2023we, designprompt2024,qu2025recycling,jiang2026when}.

A recurring limitation in existing fashion generation workflows is that they primarily emphasize visually appealing image outputs, while offering limited support for editable and structured representations that can connect user intent to downstream garment-making processes~\citep{GarmentDiffusion2025, garmentnet2025}.
For example, a request such as ``a flowy summer dress with floral patterns'' does not explicitly specify key production-relevant details such as proportions, fit and ease, or construction constraints needed for pattern generation. VLMs offer a promising way to interpret multimodal inputs and extract garment attributes, such as sleeves, necklines, and skirt types, for downstream use~\citep{chatgarment2025, fashionm32025}.
However, for everyday users, it still remains difficult to understand which attributes are controllable, such as length, looseness, or neckline depth, and how to describe desired changes in simple and effective language~\citep{styletailor2025, lilireview2025}.

Beyond early-stage ideation, an important research direction is production-oriented co-creation, in which the output is not limited to concept images but also includes artifacts that can support real garment making. This requires systems to connect user intent to representations that are both editable during interaction and compatible with downstream pattern making and manufacturing. Although some prior work has begun to connect generative models with garment representations and pattern-related pipelines~\citep{neuraltailor2022, garmentcode2023}, end-to-end workflows that translate multimodal design intent into production-oriented outputs for non-professional users remain limited.

\subsection{Sewing Pattern-Based Garment Generation}

Garment generation methods can be broadly divided into direct 3D garment generation and sewing pattern-based garment generation~\citep{luo2025FashionDesign, design2garmentcode2025}.
Direct 3D garment generation typically relies on mesh-based or differentiable representations, such as unsigned distance fields~\citep{udiff2024} and Gaussian splatting~\citep{gaussiangarment2025}, to generate garments with visually rich folds and wrinkles~\citep{wolff2023, pietroni2022}.
Although these methods can produce compelling visual results, converting arbitrary 3D garment geometry into production-oriented sewing patterns often requires additional nontrivial processing and domain expertise, which limits their applicability in manufacturing contexts.

By contrast, sewing pattern-based approaches represent garments as two-dimensional panels with stitching topology, making them more naturally aligned with production workflows~\citep{trj_review2025, luo2025FashionDesign}.
Earlier studies explored parametric pattern models for specific garment categories, such as skirts~\citep{tracy2013} and trousers ~\citep{kaixuan_pants_2019}, enabling measurement-based personalization but offering limited style diversity. GarmentCode~\citep{garmentcode2023} introduced a modular and parameterized pattern representation, which was later extended by GarmentCodeData~\citep{garmentcodedata2024} to support data-driven pattern generation. More recent systems, such as DressCode~\citep{dresscode2024} and ChatGarment~\citep{chatgarment2025}, combine generative models with language-based interfaces to improve semantic controllability from multimodal inputs. Building on these pattern-based pipelines, our work connects user design intent and body measurements to production-oriented outputs, with a particular emphasis on representations that remain editable and compatible with downstream garment construction.

\subsection{Human Body Reconstruction for Virtual Try-On}

3D human body reconstruction is a fundamental component of virtual try-on, because body shape and scale directly influence perceived fit and garment drape.
Existing approaches generally fall into two categories: scanned human models and parametric body models ~\citep{measurementstobody2025}.
Scanning can capture detailed body geometry, but it typically requires specialized hardware, tight-fitting clothing, and additional post-processing to reduce posture-related artifacts such as slouching or uneven shoulders~\citep{measurementstobody2025, pei2022, gill2015}.
These requirements limit its scalability for practical personalized try-on applications.

Parametric body models such as SMPL~\citep{smpl2015} provide low-dimensional and consistent representations of body shape and pose, and are therefore widely used in virtual try-on pipelines~\citep{dresscode2024, chatgarment2025, garmentcode2023}.
However, SMPL parameters do not directly correspond to anthropometric measurements, which has motivated substantial prior work on mapping between body measurements and parametric body representations ~\citep{li2022, xu2018, zeng20183d, allen2003, jesus3dbody2024}.
Earlier studies explored regression from measurements to low-dimensional body shape spaces~\citep{allen2003, seo2003}, although such approaches often depend on high-quality scan data. More recent methods attempt to predict measurements directly from images in order to reduce user burden~\citep{jesus3dbody2024, zeng20183d}. However, these image-based outputs can be sensitive to clothing, viewpoint, and pose, and may fail to preserve the scale consistency required for reliable fit evaluation in SMPL-based virtual try-on.

Our work focuses on improving the robustness of image-based anthropometric measurement extraction under a lightweight capture protocol that uses two user images together with user-provided height and gender for scale calibration, and then applies an existing SMPL-compatible body reshaping pipeline for virtual try-on. Rather than proposing a more accurate measurement-to-SMPL mapping method, our goal is to strengthen the reliability and scale consistency of the measurement inputs, thereby improving the trustworthiness of body-specific try-on feedback in iterative garment design.

\section{The EasyFashion System Design and Implementation} \label{sec:system}

To address \textbf{RQ1: How can a personalized fashion design system be designed for non-professional users?} We designed and implemented EasyFashion, an interactive system that supports personalized garment design through multimodal human-AI co-creation. The system is accessed through a web-based user interface and is built on an internal pipeline that transforms user inputs into structured garment specifications, personalized avatars, digital fabrics, sewing patterns, and virtual try-on. This section first presents the formative study that informed the system design, and then describes the user interface and technical implementation of EasyFashion.

\begin{table}[pos=h]
  \caption{Demographics of participants in the formative study.}
  \label{tab:pre_survey_demographics}
  \centering
  \begin{tabular}{p{0.18\textwidth} p{0.28\textwidth} p{0.38\textwidth}}
    \toprule
    Category & Non-professionals & Professionals \\
    \midrule
    Age range & 18--44 years & 25--34 years \\
    Gender ratio & Male: 6, Female: 6 & Male: 3, Female: 3 \\
    Experience with garment design & No prior experience with garment design & Over 5 years of professional experience in garment design \\
    \bottomrule
  \end{tabular}
\end{table}

\subsection{Questionnaire Survey}

This formative study was conducted to inform the design of EasyFashion by examining three questions: what difficulties non-professional users encounter in personalized garment design, how different input modalities affect design intent communication, and what conditions make an iterative human-AI co-creation workflow feel useful and trustworthy. To keep the study lightweight while still generating actionable design guidance, we combined a questionnaire survey with semi-structured interviews, allowing us to triangulate quantitative preferences with qualitative explanations. We recruited 12 non-professional participants with no prior garment design experience and 6 professional participants with more than 5 years of garment design experience, so that the study could capture both end-user needs and production-oriented considerations. Participant demographics are summarized in Table~\ref{tab:pre_survey_demographics}.

The questionnaire included single-choice, multiple-choice, and open-ended questions. It asked participants to report: (i) perceived pain points and desired support in personalized garment design; (ii) preferred ways of expressing and revising garment design intent using text, reference images, sketches, or their combinations; (iii) expectations regarding iteration efficiency and system outputs; and (iv) the perceived value of on-body visualization and virtual try-on feedback during iterative refinement. We then conducted one-on-one online semi-structured interviews to probe the reasons behind participants' survey responses. The interviews focused on past or imagined customization experiences, communication breakdowns, how participants would refine a result that is close but still incorrect, and what conditions would make AI-supported try-on feedback feel trustworthy. Throughout this section, we use anonymous identifiers N1--N12 for non-professional participants and E1--E6 for professional participants.

\begin{figure}[pos=h]
  \centering
  \includegraphics[width=\linewidth]{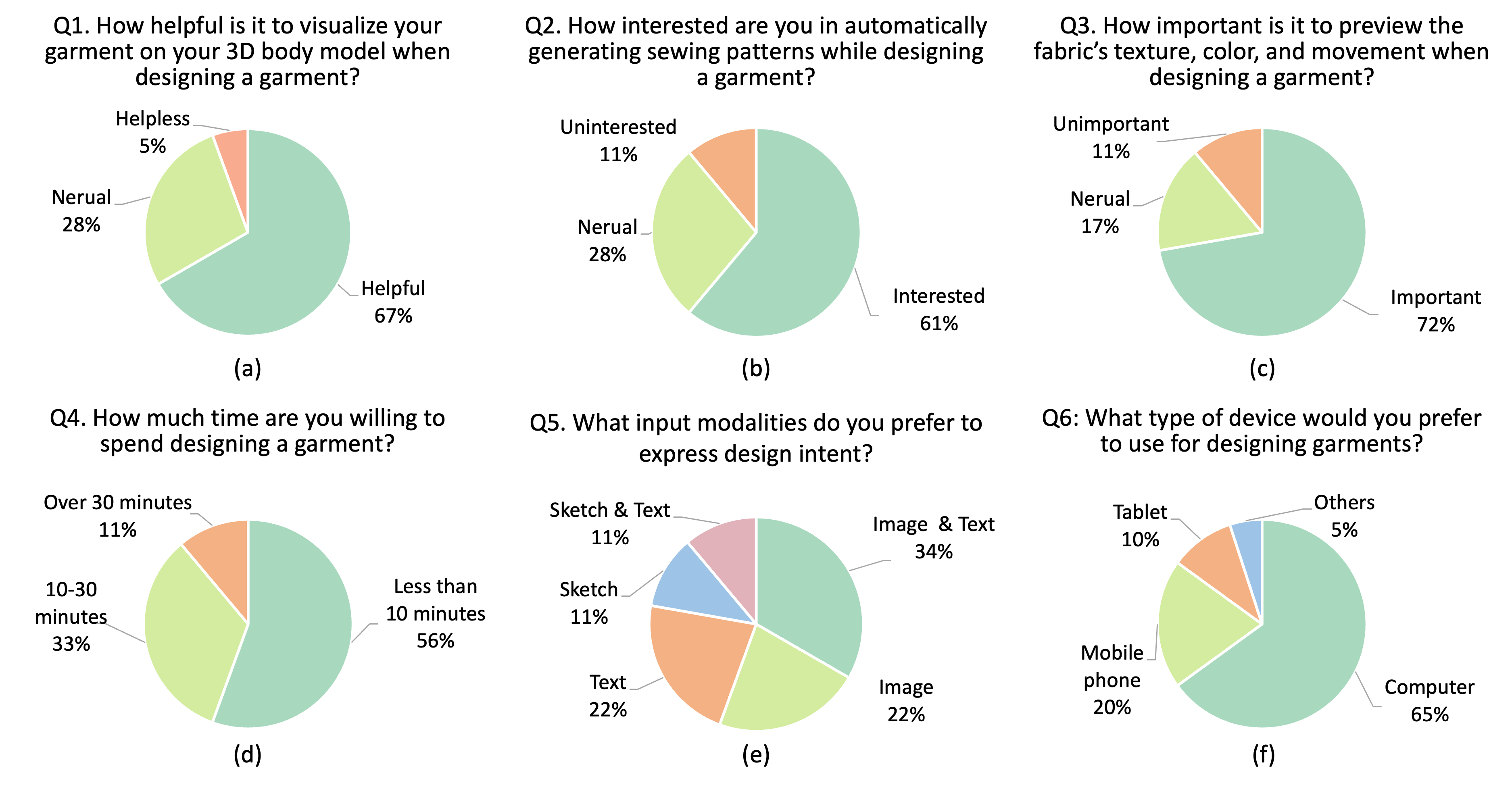}
  \caption{Insights from the pre-survey on personalized garment generation.}
  \label{fig:pre_survey}
  
\end{figure}

\subsubsection{Non-Professional Users' Perspectives}

Across interviews, non-professional participants consistently described garment customization as a high-effort, high-risk process. A major concern was the cost of repeated communication and repeated try-on, which could prolong the process and create uncertainty about whether the final garment would match the original expectation (N3, N8). Some participants associated customization mainly with formal garments such as suits and were less familiar with customization for everyday clothing. They also noted that long production cycles could lead to preference drift: by the time the garment is delivered, it may no longer feel desirable, yet the outcome is difficult to change (N4, N9). In addition, participants described social and psychological pressure when interacting with professional tailors, including concern about being seen as inexperienced, being persuaded into compromises they did not want, or not knowing whether they should insist on their original intent (N7, N10).

Survey results further revealed strong practical constraints. Specifically, 56\% of participants expected the design process to take less than 10 minutes, 33\% expected 10--30 minutes, and only 11\% were willing to spend more than 30 minutes (Fig.~\ref{fig:pre_survey} (d)). In addition, 65\% preferred using a computer rather than a mobile device for such tasks (Fig.~\ref{fig:pre_survey} (f)). Together, these findings suggest that a usable system for non-professional users should support efficient interaction and detailed inspection, and that a desktop-first workflow is appropriate for iterative refinement. 

With regard to design intent communication, non-professional users often lacked the vocabulary needed to describe construction-level details such as ease, shoulder slope, darts, or armhole shaping, and they considered freehand sketches too abstract to communicate actionable constraints (N2, N11). Instead, they frequently relied on reference images as the primary way to express intent, supplemented by short phrases indicating what to retain or modify. Survey results reflected these preferences (Fig.~\ref{fig:pre_survey} (e)): 34\% preferred reference image + text, 22\% preferred images alone, and 22\% preferred text alone, whereas fewer participants selected sketches (11\%) or sketch + text (11\%).  

When discussing the transition from concept to production, non-professional participants strongly valued try-on feedback as a way to externalize and iteratively refine conceptual design intent. They expected a workflow in which they could react to intermediate try-on results, request targeted modifications, and see these changes reflected in an editable garment representation that remained consistent with downstream production. A majority of participants wanted on-body visualization to better understand fit and appearance (Fig.~\ref{fig:pre_survey} (a)), and 72\% considered previewing fabric-related appearance such as texture and color important (Fig.~\ref{fig:pre_survey} (c)). Beyond visualization, 61\% expressed interest in having the system automatically generate production-oriented sewing patterns (Fig.~\ref{fig:pre_survey} (b)),  indicating a clear desire to move from design exploration toward physically realizable outcomes.

\subsubsection{Professional Users' Perspectives}

Professional participants provided complementary insights grounded in garment production experience. They pointed out that reference images can still be ambiguous, because the same garment may appear very different across body shapes, and fabric differences can substantially alter silhouette and drape (E1, E4). This suggests that a practical system should do more than simply accept visual references; it should also help disambiguate user intent.  

Professionals also emphasized that virtual try-on feedback must be sufficiently body-specific and scale-consistent to support decision-making (E3). Without these qualities, such feedback may still support exploration, but it remains insufficient for production-related decisions. They further stressed that try-on results should align with the generated sewing patterns and measurements, so that changes made during visual refinement correspond to changes that can actually be carried through into garment production.  

\subsubsection{Design Requirements from the Formative Study}

Based on the questionnaire and interview findings, we derived four \textbf{design requirements (DRs)} that guided the design of EasyFashion. These requirements translate user needs and production-related concerns into specific interface and system decisions.

\paragraph{\textbf{DR1:} Make design ideas easy to express.}
Non-professional users reported difficulty describing garment details using technical terminology, and many found freehand sketches too abstract to communicate clear design constraints. Instead, they more often relied on reference images and short phrases to indicate what should be kept or changed. This suggests that the system should support intent input in a way that aligns with how users naturally communicate design ideas, while minimizing the need for long prompts or specialized vocabulary. We therefore adopted reference image + text as the primary garment design input in the final interface.

\paragraph{\textbf{DR2:} Make iterative refinement easy.}
Users described customization as costly and risky when it involved repeated back-and-forth communication and long waiting cycles. They expected a workflow in which they could respond to intermediate outputs, request targeted modifications, and continue refining without restarting the process. Survey results also showed strong time constraints and a preference for computer-based interaction for detailed inspection. Together, these findings suggest that each round of refinement should be lightweight and that previous inputs and outputs should remain available across iterations. We therefore designed EasyFashion to preserve the latest inputs and outputs during regeneration and to support repeated edit-generate-evaluate cycles through a persistent garment representation. 

\paragraph{\textbf{DR3:} Provide personalized feedback for fit evaluation.}
Users valued on-body visualization because they wanted to judge fit and appearance on themselves rather than on an abstract model. Professionals further emphasized that such feedback should be body-specific and scale-consistent to support confident decisions. This suggests that virtual try-on should be grounded in a personalized body representation rather than generic visualization alone. We therefore incorporated a body reconstruction module that uses front and side images together with basic user attributes to generate a personalized avatar and estimate anthropometric measurements for try-on.

\paragraph{\textbf{DR4:} Keep design and production outputs consistent.}
Both non-professional and professional participants emphasized the importance of connecting design exploration to downstream garment making. Non-professional users expressed interest in automatically generated sewing patterns, while professionals stressed that try-on results should remain consistent with the patterns and measurements used for production. This suggests that the system should preserve an editable structured representation that links user intent, visual feedback, and production artifacts. We therefore used a structured garment specification in JSON as a persistent design state to drive both virtual try-on and sewing pattern generation and to support export for downstream communication and production.

\begin{figure}[pos=h]
  \centering
  \includegraphics[width=\linewidth]{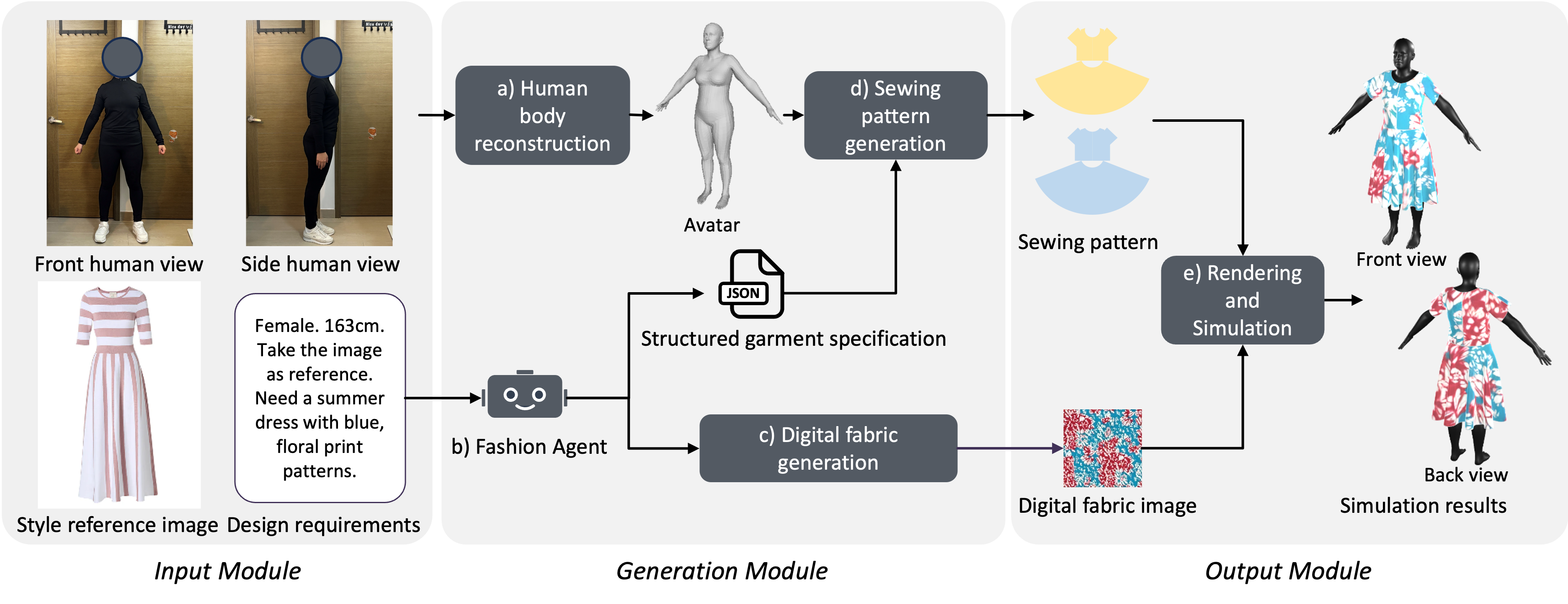}
  \caption{Internal pipeline of the EasyFashion system, consisting of three modules: the Input module (left), the generation module (middle), and the output module (right). The Input module includes Front human view, Side human view, Style reference image, and Design requirements. In the generation module, (a) Human body reconstruction: creates an Avatar from the Front human view and Side human view; (b) Fashion Agent: translates the Style reference image and Design requirements into a Structured garment specification, and additionally provides appearance guidance for Digital fabric generation; (c) Digital fabric generation: creates a Digital fabric image; and (d) Sewing pattern generation: produces a Sewing pattern from the Structured garment specification. In the output module, (e) Rendering and Simulation: combines the Sewing pattern and Digital fabric image to produce Simulation results in Front view and Back view.}
  \label{fig:pipeline}
\end{figure}

\subsection{User Interface Design}
\label{sec:ui}

EasyFashion was implemented as an interactive system with a web-based user interface and an end-to-end internal pipeline that transforms user inputs into personalized garment assets for both production and visualization, including a fitted SMPL~\citep{smpl2015} avatar, parametric sewing patterns, and virtual try-on renderings. As shown in Fig.~\ref{fig:pipeline}, the system consists of five modules: (a) human body reconstruction, (b) Fashion Agent, (c) digital fabric generation, (d) sewing pattern generation, and (e) rendering and simulation. To keep the pipeline controllable and compatible with downstream procedural pattern construction, EasyFashion uses two explicit intermediate representations: (1) a reconstructed body model with anthropometric measurements, and (2) a structured garment specification in JSON format.

EasyFashion is designed as a multimodal human-AI co-creation workflow. Although we evaluated five candidate garment-intent input conditions in our technical and user studies, the deployed interface adopts a single default setting, reference image + text, for garment design intent. In this workflow, users express design intent and revision requests through a reference image and a short text prompt, while the system maintains a persistent design state through the JSON specification and returns body-specific try-on results. This structure supports repeated cycles of editing, regeneration, and evaluation, allowing users to progressively align their design intent with production-oriented outputs.

In the final interface, reference image + text means that the garment reference image provides the primary visual cue, while the text prompt provides complementary constraints, such as collar type, sleeve style, or garment length, in order to reduce ambiguity and guide generation. Across iterations, users can update the text prompt and replace reference images directly, while the system reinterprets the revised intent into the same editable JSON representation for downstream generation.

\begin{figure}[pos=h]
  \centering
  \includegraphics[width=\linewidth]{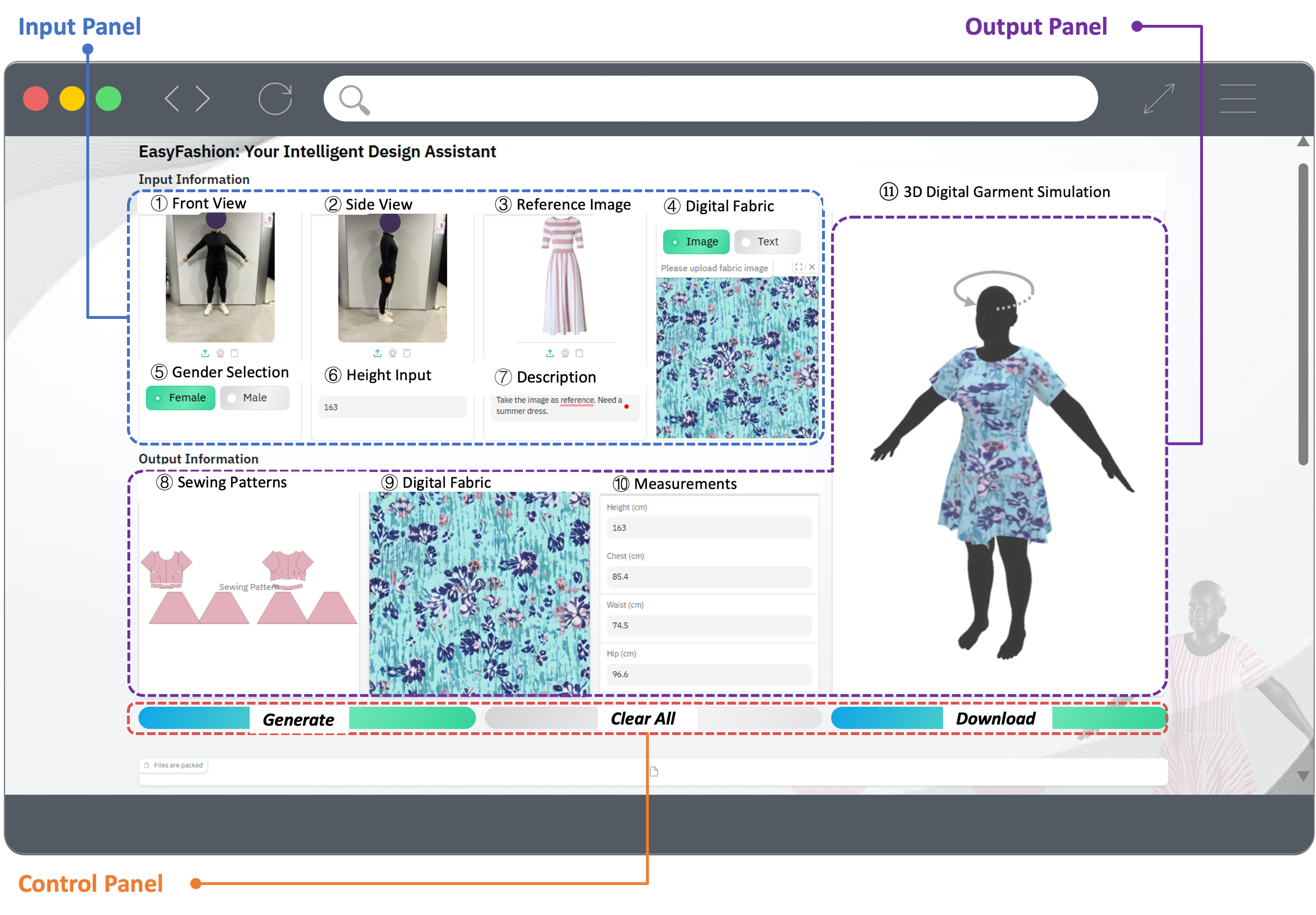}
  \caption{The web-based user interface of the EasyFashion system, consisting of Input, Output, and Control Panels.}
  \label{fig:user_interface}
\end{figure}

The web-based user interface of EasyFashion follows a three-panel layout including Input, Control, and Output, as shown in Fig.~\ref{fig:user_interface}.
This design allows users to provide body-related information and garment design intent, trigger generation iteratively, and inspect both intermediate and final outputs for further refinement. To support co-creation, the interface keeps the latest inputs and generated outputs visible throughout the refinement process, enabling users to revise their intent in context rather than restarting from scratch.

\textbf{Input panel.}
The input panel organizes the information needed for personalized garment generation by combining body-related inputs, garment design intent, and fabric appearance cues. For body personalization, users upload a front-view image and a side-view image, together with basic attributes such as gender and height. To express garment design intent, users provide a reference garment image and a short description following the reference image + text setting. The panel also supports fabric appearance specification through a fabric image and an optional text input, allowing users either to provide a target textile pattern directly or to supplement it with textual guidance.

\textbf{Output panel.}
The output panel presents both intermediate outputs and final visualization results so that users can inspect the current system output and decide whether further refinement is needed. Specifically, it displays the generated sewing patterns, the digital fabric result, the estimated body measurements, and the 3D digital garment simulation on the reconstructed avatar. By presenting these outputs in the same panel, the interface helps users ground their revisions in visible results and supports iterative co-creation.

\textbf{Control panel.}
The control panel supports the main actions required to operate the workflow, including \textit{Generate}, \textit{Clear All}, and \textit{Download}. Users click \textit{Generate} to run the pipeline after providing the required inputs, and they can repeat this process across multiple rounds as they revise images or text descriptions. To support lightweight iteration, the interface keeps previous inputs and the latest outputs available for comparison and revision. \textit{Clear All} resets the current session, while \textit{Download} exports the generated artifacts for downstream communication, further editing, or garment production.

\subsection{Technical Implementation}
\label{sec:implementation}

At the technical level, the EasyFashion system integrates an image-based body reconstruction module, a VLM-based Fashion Agent for extracting structured garment specifications, a parametric sewing pattern generator, a digital fabric generation pipeline, and a simulation-and-rendering module for virtual try-on. These components work together to support controllable garment generation and iterative use through the user interface, while maintaining explicit intermediate representations consisting of the body model with measurements and the garment JSON specification. This design supports human-AI co-creation because the system preserves an editable state that can be repeatedly updated through user revision requests and evaluated through both on-body visualization and production-oriented outputs.

\subsubsection{Human Body Reconstruction}
\label{3dhuman}

\begin{figure}[pos=h]
  \centering
  \includegraphics[width=\linewidth]{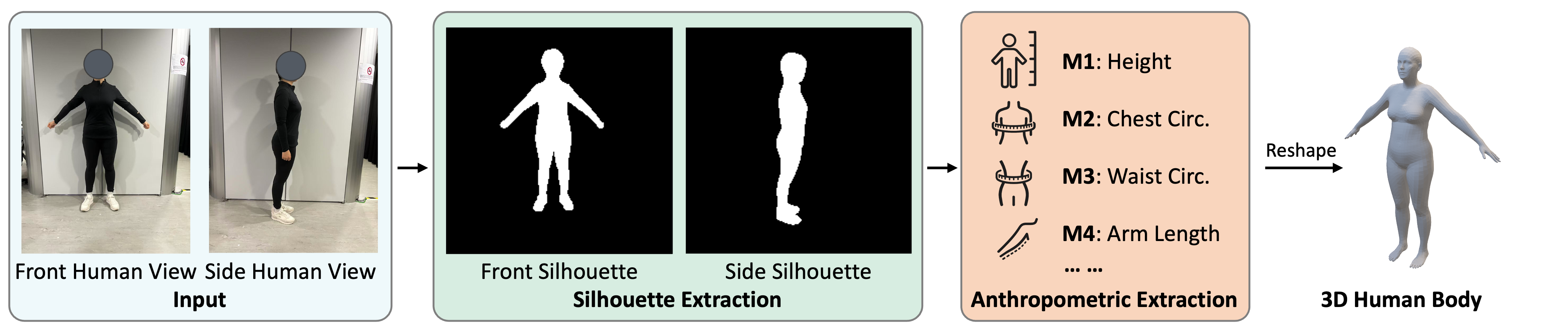}
  \caption{Pipeline of human body reconstruction, including silhouette extraction, anthropometric estimation, and body reshaping.}
  \label{fig:human_body_reconstruction}
\end{figure}

To reconstruct the user's body in a user-friendly way, we employ an image-based 3D human body reconstruction pipeline consisting of three components: \textit{Silhouette Extractor}, \textit{Anthropometric Extractor}, and \textit{Human Body Reshaper}. The pipeline takes two input images with consistent dimensions--a front view and a side view--together with basic user constraints such as height and gender, and outputs a personalized 3D human body model in SMPL~\citep{smpl2015} format and the corresponding anthropometric measurements (Fig.~\ref{fig:human_body_reconstruction}).

\paragraph{Silhouette Extractor.}
We adopt a pretrained DeepLabv3~\citep{deeplabv32017} model to segment the human body from the background and generate binary silhouettes for the front and side views. These silhouettes provide a stable representation for downstream anthropometric estimation.

\paragraph{Anthropometric Extractor.}
The Anthropometric Extractor uses the extracted front and side silhouettes together with user-provided height and gender to estimate key body measurements, including chest circumference, waist circumference, and hip circumference. Following prior work, we use a hybrid architecture with a convolutional neural network (CNN) branch to encode the silhouettes and a multilayer perceptron (MLP) branch to incorporate numerical and categorical attributes. The fused features are then mapped to an anthropometric measurement vector
\(M = \{M_1, M_2, M_3, \dots, M_n\}\),
where \(M_1\), \(M_2\), and \(M_3\) correspond to chest, waist, and hip circumference, respectively.  

\paragraph{Human Body Reshaper.}
 The Human Body Reshaper refines an SMPL body model by optimizing its shape parameters to match the predicted anthropometric measurements. SMPL is parameterized by a 10-dimensional shape vector \(\beta\) and a 72-dimensional pose vector \(\theta\)~\citep{smpl2015, isp2023}.
We optimize \(\beta\) while fixing the pose in order to obtain a body shape consistent with the predicted height and key circumferences, which are critical for garment fitting and construction~\citep{measurementstobody2025}.
The total loss is defined as:
\(L_{total} = w_{height} L_{height} + w_{chest} L_{chest} + w_{waist} L_{waist} + w_{hip} L_{hip}\),
where each feature loss is defined as
\(L_{feature} = (M_{predicted} - M_{SMPL})^2\).

Here \(M_{predicted}\) denotes the measurements estimated by the Anthropometric Extractor, and \(M_{SMPL}\) denotes the measurements computed from the current SMPL mesh. We minimize \(L_{total}\) using Adam~\citep{adam2014}, and the optimization procedure is summarized in Algorithm~\hyperref[alg:optimizer]{1}.

\begin{table}[pos=h]
  \centering
  \begin{tabular}{l}
  \hline
  \textbf{Algorithm 1: Pseudocode of the Optimization Procedure.} \\
  \hline
  \textbf{Input:} Target measurements \(M_{target} = \{\text{height}, \text{chest}, \text{waist}, \text{hip}\}\), \\
  \hspace{2em} Initial parameters \({\beta}_0\) (shape), learning rate \(\eta\), \\
  \hspace{2em} Convergence threshold \(\epsilon\), maximum iterations \(N_{max}\). \\
  \textbf{Output:} Optimized parameters \({\beta}^*\). \\
  \hline
  1. Initialize: \({\beta} \gets {\beta}_0\). \\
  2. \textbf{For} \(n \gets 1\) to \(N_{max}\) \textbf{do}: \\
  3. \hspace{2em} Compute mesh-based measurements \(M_{predicted} = f_{\text{SMPL}}({\beta})\). \\
  4. \hspace{2em} Compute loss \(L_{total} = \sum_{i} w_i \cdot (M_{{predicted}, i} - M_{{target}, i})^2\). \\
  5. \hspace{2em} \textbf{If} \(L_{total} < \epsilon\) \textbf{then break}. \\
  6. \hspace{2em} Compute gradients \(\nabla_{{\beta}} L_{total}\). \\
  7. \hspace{2em} Update: \({\beta} \gets {\beta} - \eta \cdot \nabla_{{\beta}} L_{total}\). \\
  8. \textbf{End For}. \\
  9. \textbf{Return} \({\beta}^*\). \\
  \hline
  \end{tabular}
  \label{alg:optimizer}
\end{table}

\subsubsection{Fashion Agent and Structured Garment Specification}
\label{sec:fashion_agent}

EasyFashion uses a vision-language model (VLM) to parse user garment design intent from a reference image and a short text description. The Fashion Agent produces two outputs: (1) a structured garment specification in JSON format for downstream pattern generation and 3D try-on, and (2) a text prompt used for appearance generation when the user does not provide a fabric or material input.

\paragraph{Attribute table to JSON mapping.}
Rather than requiring the VLM to output a complete JSON directly, we adopt a two-step pipeline. First, the agent predicts a compact attribute table consisting of key-value pairs that describe style-related and geometry-related decisions, such as garment category, collar or neckline type, sleeve type and length, skirt or pants type, and length- or fit-related attributes. We then map this attribute table to a predefined JSON template using deterministic rules. This design improves robustness and controllability because the attribute table is easier to normalize, while the deterministic mapping guarantees that all required fields are present for downstream parametric sewing pattern generation.

\paragraph{JSON template and validation.}
The JSON is organized under a top-level \texttt{design} key. The \texttt{design.meta} field defines the overall garment configuration, including whether the design contains an upper part, a bottom part, and a \texttt{wb} (waistband). Additional blocks describe parameters for specific garment components, such as collars, sleeves, shirts, pants, and skirt variants. 
Each parameter includes a value \texttt{v}, a type, and a valid range. Before pattern generation, the JSON is validated by type and range. Invalid numeric values are clamped to the nearest valid boundary, and unsupported categorical outputs are mapped to the closest available option, thereby ensuring compatibility with the sewing pattern generator.

\paragraph{Appearance prompt generation.}
In addition to the style-oriented JSON, the Fashion Agent generates a short text prompt for text-to-image synthesis of fabric or appearance when users do not upload a fabric input. This prompt describes appearance-related properties, such as fabric category and surface look, but remains separate from the production-facing JSON, as this format focuses on style and geometry parameters required for pattern generation.

\subsubsection{Digital Fabric Generation}
\label{sec:fabric}

EasyFashion supports two complementary paths for fabric specification: fabric image upload and text-guided fabric generation, and users may use either or both. For user-provided fabric images, the system preprocesses the input into a renderable texture through tiling and basic normalization. When no fabric image is provided, EasyFashion uses the fabric prompt generated by the Fashion Agent to drive text-guided fabric generation. We adopt a text-to-image fabric generation pipeline inspired by DressCode~\citep{dresscode2024}.
Using a fine-tuned Stable Diffusion model~\citep{stablediffusion2022,dresscode2024}, the system generates tileable diffuse maps. Following prior work~\citep{dresscode2024}, additional decoders generate corresponding roughness and normal maps from latent embeddings, enabling physically based rendering with realistic material details.

\subsubsection{Personalized Sewing Pattern Generation}
\label{sec:pattern_generation}

EasyFashion uses the structured garment JSON produced by the Fashion Agent (Sec.~\ref{sec:fashion_agent}) as the input to GarmentCode~\citep{garmentcode2023,garmentcodedata2024} for parametric sewing pattern generation. GarmentCode provides modular pattern templates that can be instantiated and parameterized according to garment type and style. We combine style attributes from the JSON with user-specific anthropometric measurements (Sec.~\ref{3dhuman}) from the body reconstruction stage to generate sewing patterns customized to the reconstructed body shape.

Because the sewing patterns are derived from the same editable JSON and the personalized measurements, the system can regenerate patterns after specification updates while maintaining a consistent link among: (i) the current design state, (ii) the 3D virtual try-on, and (iii) the production-oriented pattern outputs. The output includes 2D pattern panels and stitching metadata needed for downstream garment assembly, simulation, and export.

\subsubsection{Rendering and Simulation}
\label{sec:render_sim}

For visualization, the system assembles the generated sewing patterns and performs garment simulation on the reconstructed SMPL avatar. We use a customized pipeline based on the NVIDIA Wrap tool~\citep{warp2022, garmentcodedata2024} to perform virtual stitching and garment simulation. Finally, the system renders 3D virtual try-on results using physically based materials derived from either the uploaded fabric image or the generated fabric maps. All generated artifacts can be exported through the user interface for downstream communication and garment production.

\section{Technical Evaluation}
\label{sec:evaluation}

To address \textbf{RQ2: Is EasyFashion effective for supporting personalized fashion design by non-professional users?} We conducted a set of technical evaluations to assess the system across its core components. Specifically, the evaluation focuses on four aspects: the accuracy of human body reconstruction, the robustness of the Fashion Agent under different input modality conditions, the system's generalization capability across diverse body shapes and garment styles, and the comparative advantage of personalized garment generation over a standard mass-customization approach. Based on these goals, we formulated the following hypotheses:

\begin{itemize}
    \item \textbf{H1}: EasyFashion can accurately estimate body measurements and reconstruct plausible 3D body shapes.
    \item \textbf{H2}: Multimodal inputs enable the Fashion Agent to produce more accurate garment specifications than single-modality inputs.
    \item \textbf{H3}: EasyFashion can generalize across diverse body shapes and garment styles.
    \item \textbf{H4}: EasyFashion can produce better-fitting personalized garments than a standard sizing system.
\end{itemize}

\subsection{Performance Evaluation of Human Body Reconstruction}
\label{eval_human_body}

This section evaluates the image-based human body reconstruction pipeline described in Section~\ref{3dhuman}. Because body-specific virtual try-on is a key part of the EasyFashion workflow, this evaluation examines whether the reconstructed bodies provide sufficiently accurate anthropometric measurements and visually plausible body shapes for downstream fit-related visualization. We therefore conduct both qualitative and quantitative comparisons against scan-derived ground truth. 

\subsubsection{Dataset Preparation}

To construct the test dataset, we recruited 42 participants, including 14 females and 28 males. This anthropometric survey was approved by the Departmental Research Committee of the relevant institution, and informed consent was obtained from all participants. All procedures were conducted in accordance with applicable ethical guidelines.

During data collection, participants wore tight-fitting clothing or no upper clothing in order to reduce the influence of apparel on scan-derived body measurements. Body scans were acquired using ANTHROSCAN~\footnote{\url{https://www.humaneticsgroup.com/products/body-scanning-solutions/anthroscan-software}}, which supports automated extraction of anthropometric measurements and anatomical landmarks according to standardized measurement protocols. The scan-reported anthropometric measurements were used as ground truth. After scanning, front-view and side-view images of each participant were also captured. The raw 3D scan data were manually cleaned to remove noise and artifacts, and the extracted anthropometric dimensions were used for subsequent evaluation.

To ensure a fair comparison between predicted measurements and scan-derived ground truth, we adopted the same circumference definitions as those used by the scanning software. Specifically, chest was defined as the maximum torso circumference in a standing pose, waist as the minimum upper-body torso circumference, and hip as the maximum lower-body circumference. All predicted measurements were aligned with the corresponding scan-reported dimensions under these definitions.

\subsubsection{Qualitative Evaluation} 
\begin{figure}[pos=h]
  \centering
  \includegraphics[width=\linewidth]{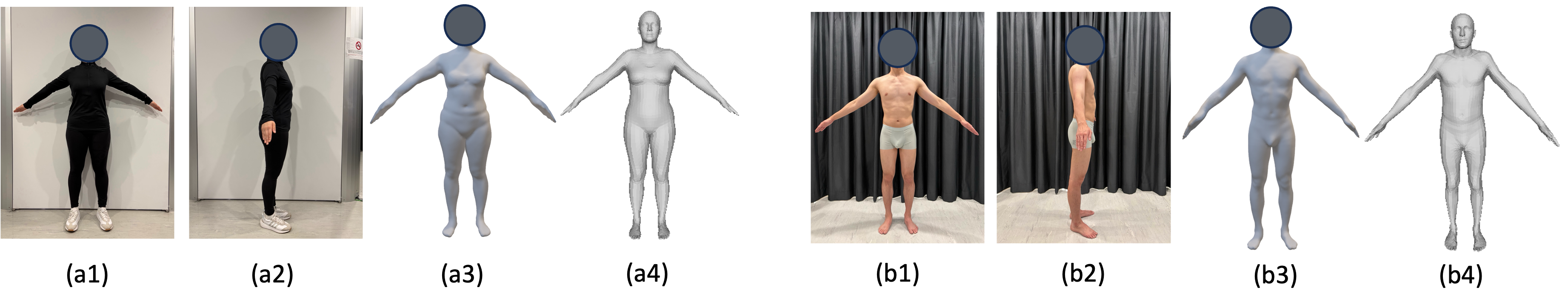}
  \caption{Examples of scanned and reshaped bodies. (a1--4) show a female subject: (a1) front-view image, (a2) side-view image, (a3) scanned body, and (a4) reshaped body.
  (b1--4) show a male subject: (b1) front-view image, (b2) side-view image, (b3) scanned body, and (b4) reshaped body.}
  \label{fig:body_reconstruction}
\end{figure}

Fig.~\ref{fig:body_reconstruction} presents representative qualitative comparisons between the scanned bodies and the reconstructed bodies generated by our method. For each example, the figure includes the front-view image, side-view image, scanned 3D body, and reconstructed 3D body. Across both female and male examples, the reconstructed bodies are visually consistent with the corresponding scanned body shapes. These results suggest that the proposed pipeline can recover plausible 3D body geometry from lightweight image-based inputs.

\begin{table}[pos=h]
  \caption{Body measurement accuracy by gender (percentage; mean \(\pm\) std).}
  \label{tab:quantitative_results}
  \centering
  \begin{tabular}{lcccc}
    \toprule
     & Height Accuracy & Chest Accuracy & Waist Accuracy & Hip Accuracy\\
    \midrule
    Female & 99.17 (\(\pm 0.01\)) & 91.55 (\(\pm 0.06\)) & 91.62 (\(\pm 0.06\)) & 95.18 (\(\pm 0.04\)) \\
    Male   & 98.91 (\(\pm 0.15\)) & 90.95 (\(\pm 0.05\)) & 85.60 (\(\pm 0.08\)) & 96.63 (\(\pm 0.03\)) \\
    \bottomrule
  \end{tabular}
\end{table}

\subsubsection{Quantitative Evaluation} 
We evaluated measurement accuracy using the following definition: \(\mathrm{Accuracy} = 1 - \frac{|\hat{y}-y|}{y}\), where \(y\) denotes the scan-derived ground-truth measurement and \(\hat{y}\) denotes the predicted measurement, in centimeters. 
As summarized in Table~\ref{tab:quantitative_results}, the reconstruction pipeline achieves high overall accuracy across both genders. For female participants, the average accuracies are 99.17\% for height, 91.55\% for chest, 91.62\% for waist, and 95.18\% for hip. For male participants, the corresponding accuracies are 98.91\% for height, 90.95\% for chest, 85.60\% for waist, and 96.63\% for hip. Overall, height and hip measurements show consistently high accuracy with relatively low variability, whereas chest and especially waist are more challenging, with male waist estimation showing the lowest performance among the reported measures. Nevertheless, these quantitative results are broadly consistent with the qualitative observations in Fig.~\ref{fig:body_reconstruction}, suggesting that the reconstructed bodies are suitable for downstream applications such as virtual try-on.

\subsection{Performance Evaluation of Inputs for the Fashion Agent}
\label{eval_vlm}

This section evaluates how effectively the Fashion Agent interprets garment design intent under different input modality conditions. As described in Section~\ref{sec:fashion_agent}, the Fashion Agent combines a vision-language model with prompt engineering to produce a structured garment specification in JSON format. In this experiment, sewing pattern generation and rendering are not treated as the end goal in themselves; instead, they serve as verifiable downstream evidence of whether the Fashion Agent has correctly understood the user's design intent. If the Agent interprets the input correctly, it should produce a structured JSON that can be consistently translated into a plausible sewing pattern and a corresponding try-on result. Accordingly, we evaluate the Agent from two perspectives: structural correctness of the generated sewing patterns and visual consistency of the rendered results with the intended design. The VLM used in this study is Gemini-2.5-Flash~\citep{gemini2025}.

\subsubsection{Dataset Preparation}
\label{sec:eval_dataset}

The dataset for this experiment is a subset of the GarmentCode dataset~\citep{garmentcode2023,garmentcodedata2024}, selected to cover a broad range of common garment categories and styles. The final subset contains 64 samples, including 8 pants, 12 tops, 6 hoodies, 22 dresses, 8 sets, and 8 skirts. To support consistent evaluation of the Fashion Agent's interpretation, we collaborated with an experienced fashion designer with more than 10 years of professional experience to manually sketch the ground-truth sewing patterns for each garment and provide textual descriptions of its key design features. In addition, GPT-4o was used to generate reference images of people wearing garments corresponding to the sewing patterns.

Based on the input modality preferences identified in the formative study (Fig.~\ref{fig:pre_survey} (e)), we compared five design-intent input conditions: (1) reference image + text, (2) reference image-only, (3) sketch + text, (4) sketch-only, and (5) text-only. Given one of these inputs, the Fashion Agent produces a structured JSON encoding garment style, structure, and design elements. This JSON is then passed to the sewing pattern generation and rendering modules to produce sewing patterns and try-on visualizations. By comparing outputs across the five conditions, we assess which modality combination most reliably supports accurate structured understanding of garment design intent.

\begin{figure}[pos=h]
  \centering
  \includegraphics[width=\linewidth]{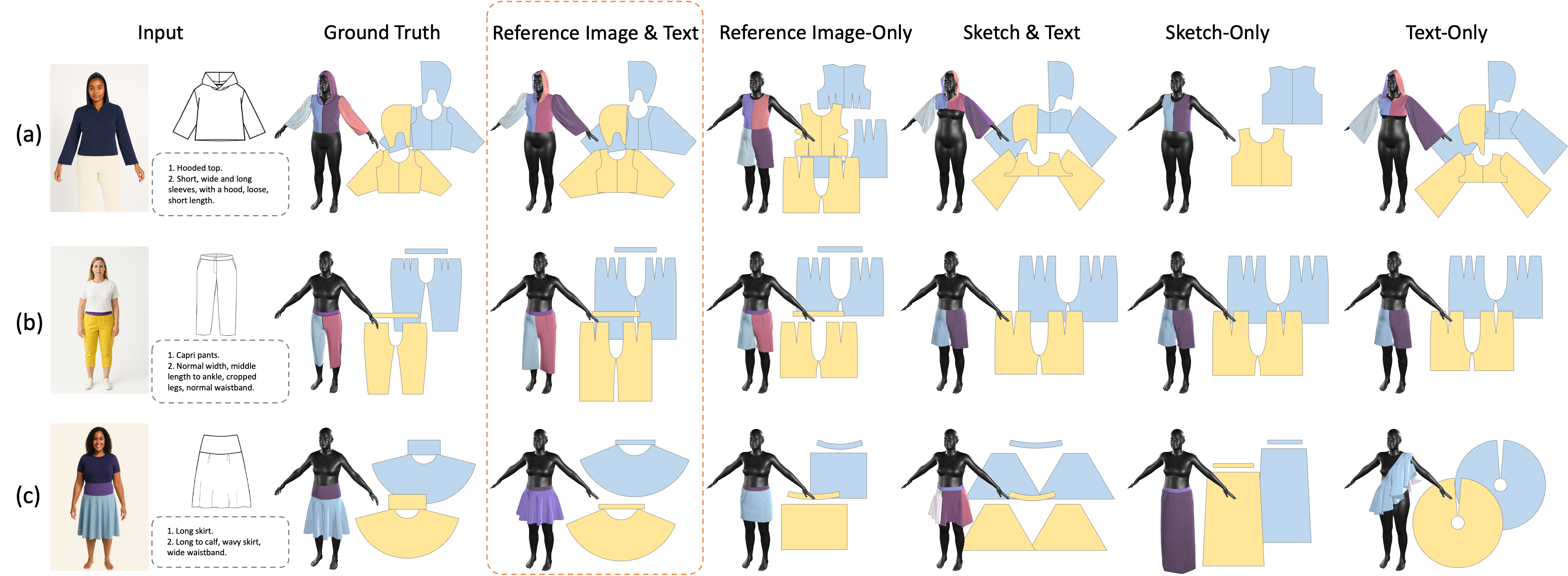}
  \caption{Comparison of generated results under five input conditions (reference image + text, reference image-only, sketch + text, sketch-only, and text-only) across three garment cases with different styles: (a) hoodie, (b) pants, and (c) skirt. For each case, the reference image + text condition yields results most closely aligned with the ground truth in terms of both garment appearance and sewing pattern structure. }
  \label{fig:sp_examples}
\end{figure}

\subsubsection{Qualitative Evaluation}
\label{sec:eval_qualitative}

Fig.~\ref{fig:sp_examples} presents representative qualitative comparisons across the five input conditions. Overall, reference image + text most consistently produces outputs whose sewing pattern structures and rendered appearances align with the intended design.

For the hoodie example, \textit{reference image + text}, \textit{sketch + text}, and \textit{text-only} all generate hoodie-like outputs, but only \textit{reference image + text} captures the intended garment length accurately. \textit{Sketch + text} and \textit{text-only} deviate in proportion, while \textit{reference image-only} and \textit{sketch-only} fail to recover the correct hoodie structure. For the capri pants example, all five conditions correctly capture the general garment category, but \textit{reference image + text} most accurately reproduces the intended length and waistband detail, whereas the remaining conditions show omissions or proportion errors. For the dress example, all five conditions produce dress-like garments, but fine-grained design elements differ substantially. \textit{reference image + text} is closest to the ground truth in both overall structure and length. \textit{Reference image-only} fails to recover the expected wavy skirt, and \textit{sketch-only} produces an overly long silhouette. Although \textit{text-only} can sometimes encode a wavy skirt in the sewing pattern, dimensional inconsistencies may prevent successful try-on rendering, indicating a mismatch between the Agent's structured output and physically plausible pattern parameters. Taken together, these examples suggest that \textit{reference image + text} provides the strongest grounding for the Fashion Agent to generate structured specifications that downstream modules can reliably realize.

\subsubsection{Quantitative Evaluation}
\begin{figure}[pos=h]
    \centering
    
    \includegraphics[width=0.9\linewidth]{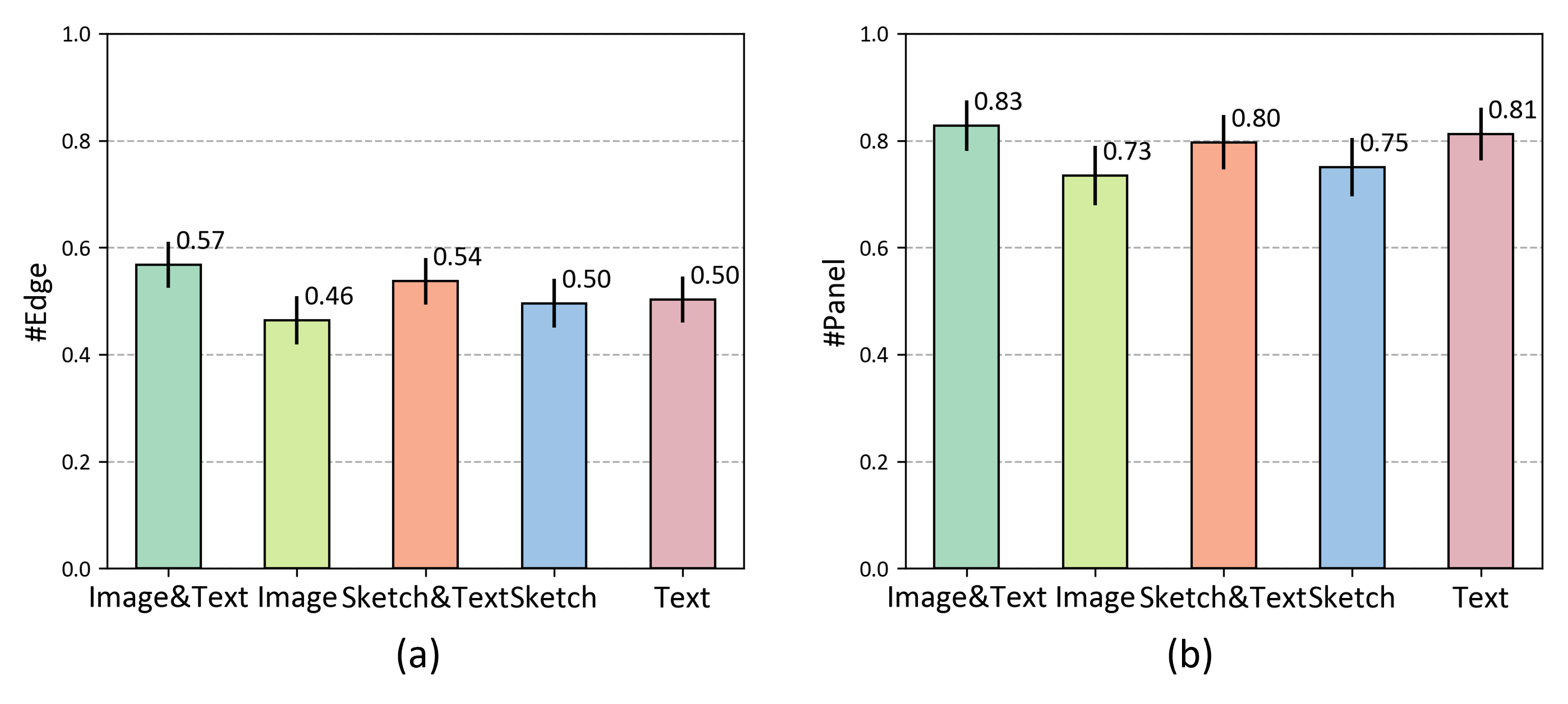}
    \caption{Quantitative comparison of edge accuracy and panel accuracy of generated sewing patterns across five input conditions.}
    \label{fig:sp_barchart}
\end{figure}

In addition to qualitative inspection, we evaluated the structural correctness of the sewing patterns derived from the Fashion Agent's JSON outputs. Following prior work~\citep{neuraltailor2022}, we compared the number of panels and the number of edges in each generated pattern against the ground-truth pattern (Fig.~\ref{fig:sp_barchart}), and report \emph{panel accuracy} and \emph{edge accuracy}.

\textit{Panel accuracy.}
Panel accuracy measures the fraction of garments whose predicted panel count exactly matches the ground truth:
$\mathrm{Acc}_{\text{panels}}=\frac{1}{N}\sum_{i=1}^{N} c_i$, where $c_i=\mathbf{1}[K_i=\hat{K}_i]$.
Here, $\mathbf{1}[\cdot]$ is the indicator function.
$N$ is the number of test garments, and $K_i$ and $\hat{K}_i$ denote the ground-truth and predicted number of panels for garment $i$, respectively.

\textit{Edge accuracy.} Edge accuracy measures, for each garment, the average fraction of panels whose edge count is predicted correctly:
$\mathrm{Acc}_{\text{edges}}=\frac{1}{N}\sum_{i=1}^{N}\left(\frac{1}{K_i}\sum_{j=1}^{K_i} e_{i,j}\right)$, where $e_{i,j}=\mathbf{1}[E_{i,j}=\hat{E}_{i,j}]$.
Here, $E_{i,j}$ and $\hat{E}_{i,j}$ are the ground-truth and predicted number of edges for the $j$-th panel in garment $i$, respectively.

As shown in Fig.~\ref{fig:sp_barchart}, reference image + text achieves the best overall performance, with a panel accuracy of 0.83 and an edge accuracy of 0.57. This indicates that the combination of visual reference and textual constraint most reliably enables the Fashion Agent to produce structured specifications consistent with the ground-truth pattern structure. Sketch + text also performs well, although slightly worse, likely because sketches are more abstract and provide fewer explicit geometric cues than reference garment images. 
Among the single-modality conditions, text-only performs best, suggesting that explicit semantic descriptions remain useful when visual grounding is absent. By contrast, reference image-only performs worse despite its rich visual content, which is consistent with the observation that inferring garment structure from images alone remains difficult for VLM-based agents without explicit semantic constraints. Overall, these results support \textbf{H2} and show that multimodal inputs outperform single-modality inputs, with reference image + text providing the strongest complementary support for both semantic intent and structural grounding.

\subsection{System Generalization Capability}
\begin{figure}[pos=h]
  \centering
  \includegraphics[width=\linewidth,height=.78\textheight,keepaspectratio]{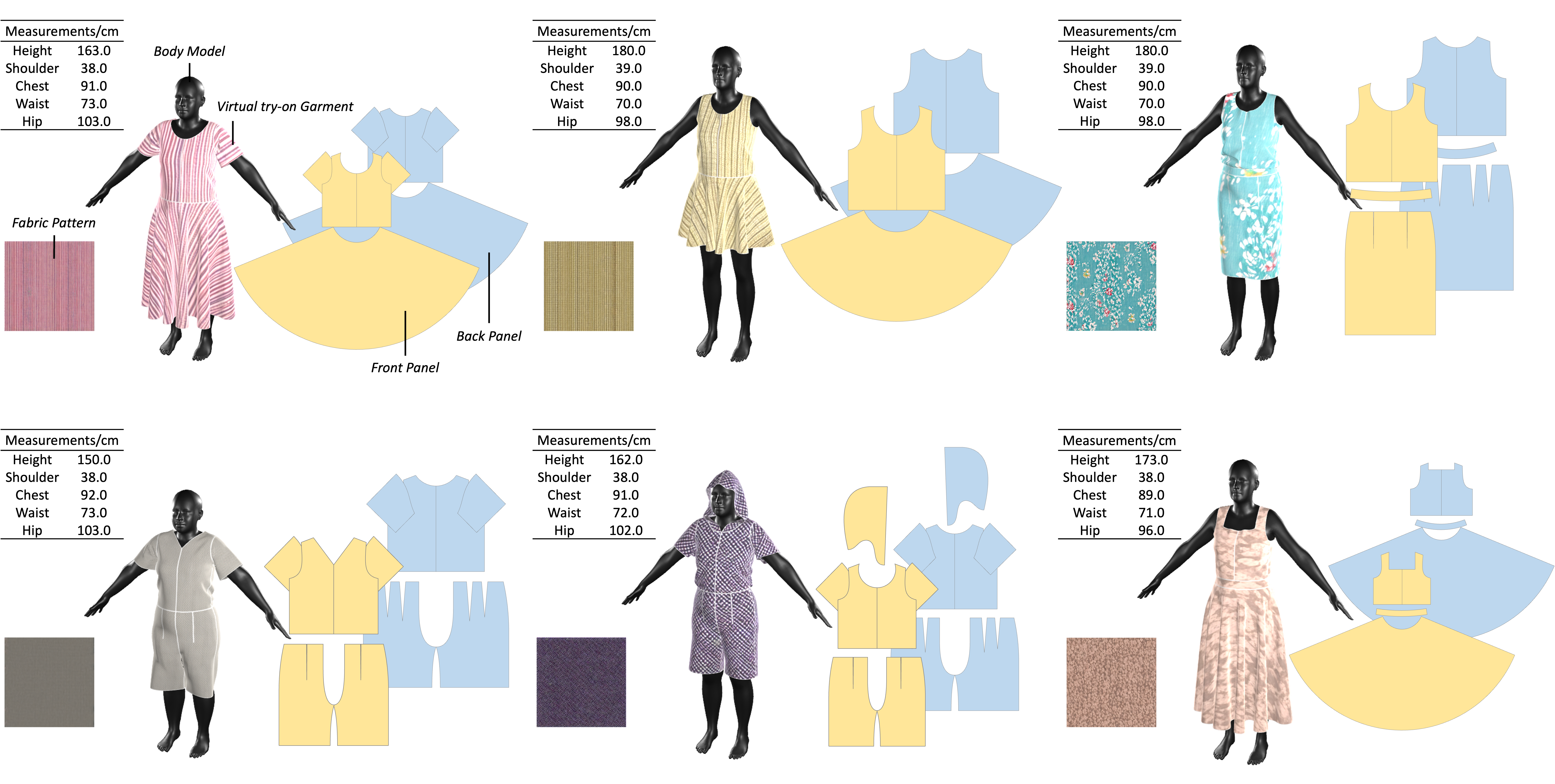}
  \caption{Different examples showcasing EasyFashion's application across participants with diverse body shapes and a variety of garment styles. (Unit: cm)}
  \label{fig:more_examples}
\end{figure}

To examine \textbf{H3}, we evaluated whether EasyFashion can generalize across a diverse range of body shapes and garment styles, rather than only reproducing a narrow set of curated examples. Fig.~\ref{fig:more_examples} presents additional outputs generated for participants with diverse body proportions and a variety of garment categories and style preferences. Across these examples, the system produces outputs that remain visually coherent and aesthetically appropriate while adapting to differences in body shape and garment type. Although this analysis is qualitative and preliminary, it provides evidence that EasyFashion can generalize across variations in body proportions, personal style preferences, and garment categories.  

\subsection{Design Method Comparison}
\begin{figure}[pos=h]
  \centering
  \includegraphics[width=\linewidth]{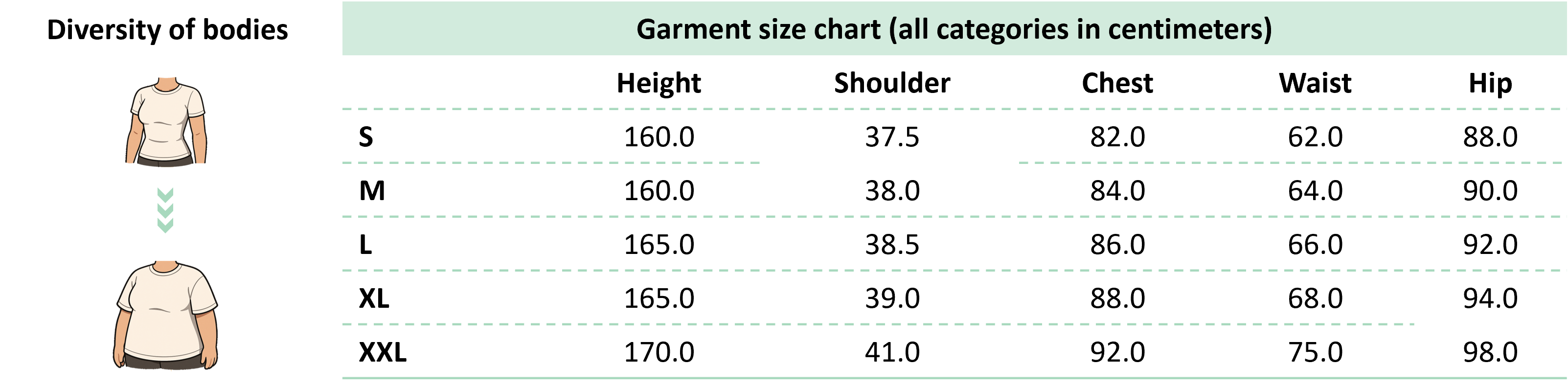}
  \caption{Standard size reference adapted from the Chinese women's ready-to-wear size designation standard (GB/T 1335.2-2008)~\citep{women_size_standard2008}.}
  \label{fig:size_chart}
\end{figure}

To evaluate \textbf{H4}, we compared garments generated by EasyFashion with garments produced under a standard sizing system or mass-customization approach. For this comparison, we constructed the garment size chart in Fig.~\ref{fig:size_chart} by adapting the Chinese women's ready-to-wear size designation standard (GB/T 1335.2-2008)~\citep{women_size_standard2008} to the garment categories used in our study. Standard size systems compress a continuous distribution of body shapes into a small number of discrete size labels, such as S to XXL. As a result, many users are forced to choose a size that does not adequately match their body proportions, leading to compromised fit.

\begin{figure}[pos=h]
  \centering
  \includegraphics[width=\linewidth,height=.78\textheight,keepaspectratio]{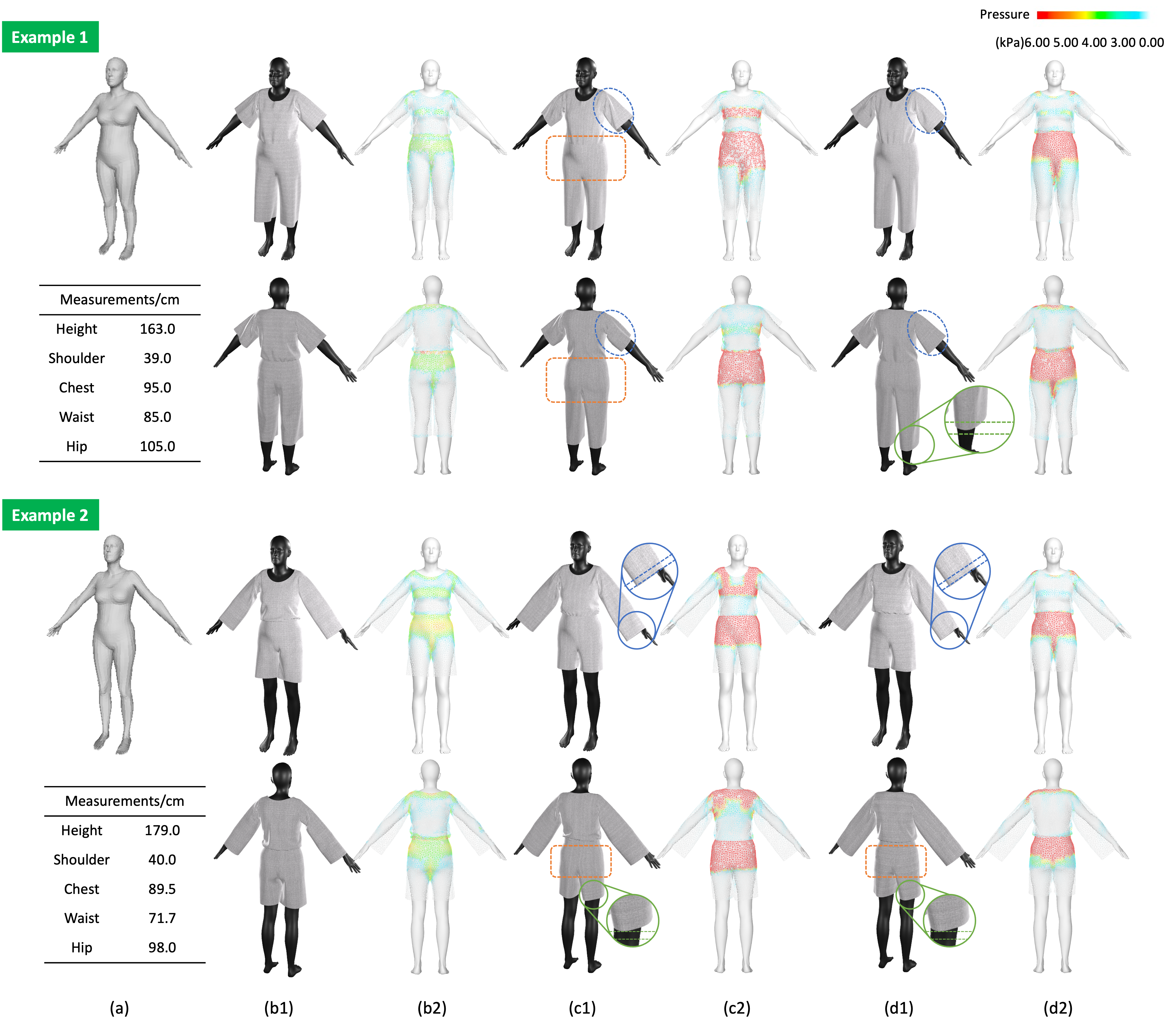}
  \caption{Comparison of garment simulation and pressure maps for two distinct body shapes under custom and standard sizes. Column (a) shows the body models and key measurements. Columns (b1), (c1), and (d1) show the simulated garments generated by our system in the custom size, standard size XL, and standard size XXL, respectively. Columns (b2), (c2), and (d2) show the corresponding pressure maps for these three size conditions.}
  \label{fig:sizes_example}
\end{figure}

Fig.~\ref{fig:sizes_example} illustrates this issue using two representative body types that are not well served by standardized sizing. For each case, we compare the personalized garment generated by EasyFashion with garments generated in common standard sizes, specifically XL and XXL. In Example 1, the user typically purchases XL or XXL in the Chinese market, yet still experiences incompatible trade-offs between hip fit and garment length. The personalized garment reduces this mismatch by jointly satisfying width-related and length-related requirements through measurement-conditioned generation. The corresponding pressure maps (Fig.~\ref{fig:sizes_example} (b2), (c2) and (d2)) further show that, compared with the standard-size garments, the personalized result exhibits a more balanced pressure distribution with fewer localized high-pressure regions, suggesting improved comfort and reduced strain in problematic areas.

In Example 2, the user's primary difficulty is insufficient garment length.
Choosing larger ready-to-wear sizes may still fail to meet the desired pant or sleeve length, while introducing tightness or silhouette distortion elsewhere.
Personalization directly addresses this limitation by generating patterns that follow the user's body proportions rather than the assumptions embedded in the size chart.
Consistent with this, the pressure maps indicate that improvements are not only visual, since length is improved, but also relate to physical comfort through reduced tightness.
This reinforces that measurement-driven generation can better satisfy both comfort-related and appearance-related needs.

\section{User Evaluation}
\label{sec:user_study}

To address \textbf{RQ3: Is EasyFashion user-friendly for supporting personalized fashion design by non-professional users?} We conducted a user evaluation to examine subjective experience across five dimensions: system usability, perceived ease of use, information quality, perceived authenticity, and overall satisfaction. We also included a production case study to examine whether the system can support the transition from design intent to real garment making in practice. Based on these goals, we formulated the following hypotheses:
\begin{itemize}
    \item \textbf{H5}: Users perceive EasyFashion as easy to use for expressing fashion design ideas.
    \item \textbf{H6}: Users can reach a satisfactory garment design within a small number of refinement rounds.
    \item \textbf{H7}: EasyFashion can support users in progressing from design ideas to production-oriented garment outputs.
\end{itemize}

Following the technical evaluation, we conducted a user study to assess the usability and interaction experience of EasyFashion for non-professional users. The study consisted of two parts.
\textit{Study A} compared five input conditions in order to identify the most effective interaction setting for the final system.
\textit{Study B} then evaluated this selected setting with the same participants through iterative use, combining quantitative ratings with qualitative feedback from a brief post-task interview. Across both studies, we conceptualize interaction with EasyFashion as a form of human-AI co-creation, in which participants iteratively expressed design intent and revision requests while the system regenerated outputs and maintained a consistent link between body-specific virtual try-on feedback and production-oriented artifacts.

\subsection{Participants}
\label{sec:user_participants}

\begin{figure}[pos=h]
  \centering
  \includegraphics[width=\linewidth]{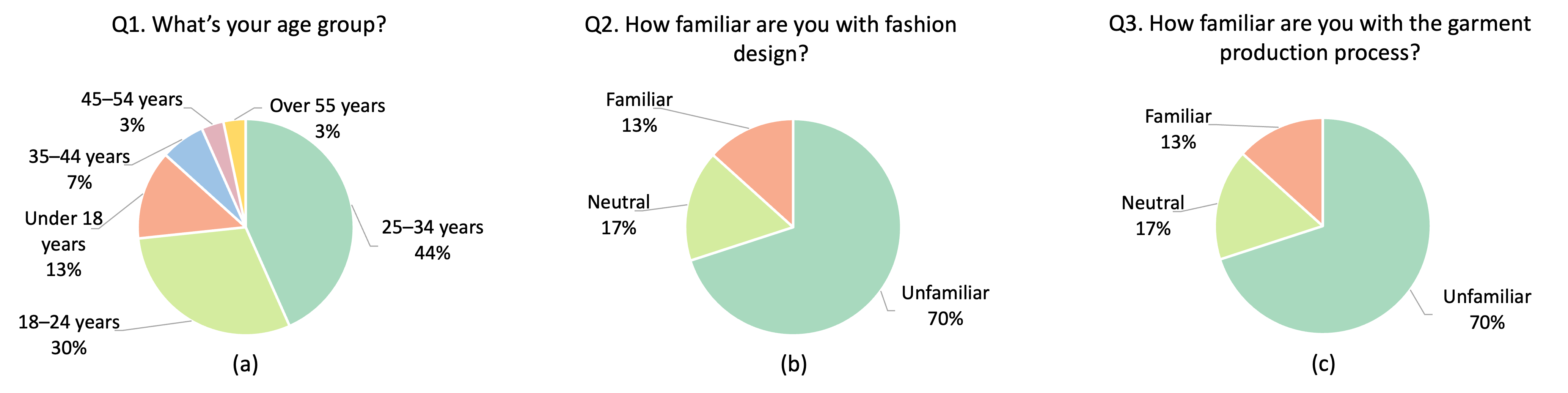}
  \caption{Participant demographics and prior experience with fashion design and garment production in the user study.}
  \label{fig:user_demographics}
\end{figure}

We recruited 30 participants (15 female and 15 male) from online forums. Participants ranged in age from 18 to 55 years, with the largest groups being 25--34 years old (44\%) and 18--24 years old (30\%). Most participants reported no prior experience with garment design or production (70\%), while 13\% reported relevant experience and 17\% selected a neutral response option (Fig.~\ref{fig:user_demographics}).
In the qualitative excerpts reported below, we refer to participants as P1--P30.

\subsection{Experimental Design}
\label{sec:Experimenta_Design}

We evaluated EasyFashion through two sequential user studies, each designed to address a distinct aspect of interaction design and practical usability. Below, we describe the independent variables, dependent variables, and measurement methods for each study.

\textit{\textbf{Study A:} Input Condition Comparison.}
The independent variable was input modality. We employed a within-subjects design with five conditions: 
(1) reference image + text, 
(2) reference image-only, 
(3) sketch + text, 
(4) sketch-only, and 
(5) text-only. %
Participant perceptions were measured using a 25-item questionnaire adapted from established instruments. The core items were based on the Post-Study System Usability Questionnaire (PSSUQ)~\citep{pssuq1992},  supplemented with items from the System Usability Scale (SUS)~\citep{sus1996} and the Object Authenticity (OBAU) scale~\citep{muralart2025}. All items used a 7-point Likert scale, where 1 = Strongly Agree / most positive and 7 = Strongly Disagree / most negative. Responses were aggregated into two subscales for analysis: System Usability (SYSUSE) and Perceived Ease of Use (PEU). These measures allowed direct comparison of interaction quality across the five input conditions.

\textit{\textbf{Study B:} Selected Setting Evaluation.}
The independent variable was fixed to the interaction setting that received the best overall usability results in \textit{Study A}, namely, reference image + text. We used the same 25-item questionnaire to evaluate the final system in a more open-ended, iterative task. Responses were aggregated into five subscales: System Usability (SYSUSE), Perceived Ease of Use (PEU), Information Quality (INFOQUAL), Object Authenticity (OBAU), and Overall Satisfaction. For all subscales, lower scores indicate more favorable ratings. This design enabled us to evaluate the selected interaction setting in a more realistic usage scenario that more closely reflects how non-professional users would engage with the system in practice.

\subsection{User Experimental Procedure}
\label{sec:user_procedure}
After providing informed consent and completing a demographic questionnaire, participants received a researcher-guided tutorial and then used the system. We conducted Study A and Study B sequentially, with each study serving a different evaluation purpose.

\begin{figure}[pos=h]
    \centering
    \includegraphics[width=\linewidth]{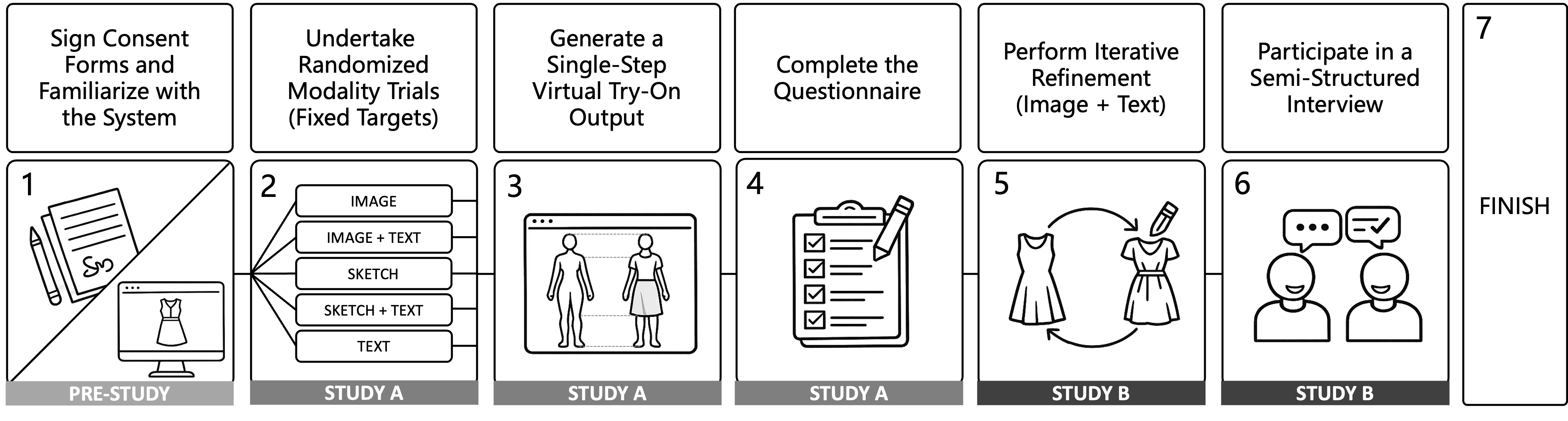}
    \caption{User experimental procedure for the two-stage study workflow, including the input condition comparison in Study A and the selected setting evaluation in Study B.}
    \label{fig:placeholder}
\end{figure}

\textit{\textbf{Study A:} Input Condition Comparison.}

To reduce ordering effects, the five conditions were presented in a randomized order for each participant. To ensure that all participants worked with consistent garment targets, the reference images and sketches were provided by the researchers. For text-based conditions, participants entered their own descriptions to specify garment details. For body-related inputs, participants could either upload their own body images or use sample images provided by the researchers.

Within each condition, participants generated a try-on result for the same target garment, inspected the output, and then completed the questionnaire. To reduce confounding effects from iterative trial-and-error, participants were instructed to perform only one generation per condition, and no repeated regeneration or prompt revision was allowed within that condition. This design isolates how effectively each modality supports one-step design intent expression and interpretation.  

\textit{\textbf{Study B:} Selected Setting Evaluation.}
After all participants completed \textit{Study A}, we aggregated the ratings to identify the interaction setting used in the final system, which was reference image + text. Participants were not shown the aggregated results from \textit{Study A}. They then evaluated this selected setting by using the system more freely to generate garments they personally wanted through iterative refinement. Participants could revise both their text prompts and reference images between rounds while keeping prior inputs as contextual support, and they stopped when they felt the output reasonably matched their intent.

To characterize co-creation behavior, we recorded the number of generation rounds per participant in \textit{Study B} and collected self-reported reasons for stopping during the post-task interview. In practice, participants typically reached a satisfactory garment within 2 to 3 rounds and rarely exceeded five rounds. The most common stopping reason was that the output was already close enough to their intent, while a secondary reason was reduced motivation, since the task was framed as part of a study rather than a real purchasing or production decision. All experiments were conducted on a server with an Intel Xeon E5-2673 V3 CPU, 32 GB RAM, and an NVIDIA GeForce RTX 3060 Laptop GPU. The end-to-end generation time was approximately 117 seconds per run, including rendering and simulation.

\subsection{Data Analysis}
\label{sec:Data_analysis}

We collected both quantitative and qualitative feedback through the standardized questionnaire and semi-structured interviews. All quantitative ratings were measured on a 7-point Likert scale. Qualitative insights were obtained from open-ended questionnaire responses and brief post-task interviews.

For \textit{Study A}, we focused on SYSUSE and PEU in order to compare interaction quality across the five input conditions. Because each participant rated all five conditions and the responses were ordinal, we used a Friedman test~\citep{friedman1937use}, followed by Holm-corrected post-hoc pairwise comparisons~\citep{holm1979simple}.

For \textit{Study B}, we report descriptive statistics--including the mean and standard deviation--for each metric under the selected interaction setting, where lower values indicate better ratings.

Qualitative findings were drawn from the post-task interviews. Each interview lasted approximately 3 minutes and focused on participants' overall impressions, perceived strengths and limitations of the system, experiences with iterative refinement, and suggestions for improvement. Interview responses were transcribed for analysis.

\subsection{Experimental Results}
\label{sec:user_results}

\begin{figure}[pos=h]
  \centering
  \includegraphics[width=\linewidth]{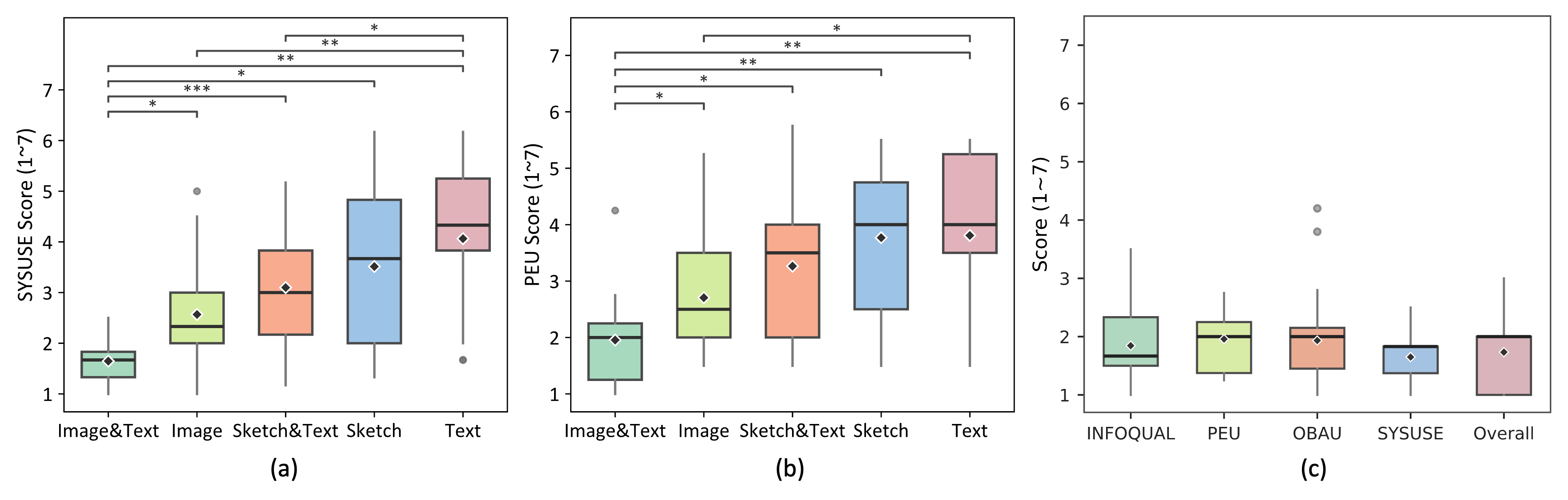}
  \caption{User experience evaluation and comparisons across Study A and Study B.
  (a--b) Study A: comparison of five input conditions in terms of (a) SYSUSE and (b) PEU.
  (c) Study B: results for the selected setting across five metrics (lower is better).
  Asterisks indicate significant differences (* $=p<0.05$; ** $=p<0.01$; *** $=p<0.001$).}
  \label{fig:input_evaluation}
\end{figure}

Fig.~\ref{fig:input_evaluation} (a--b) summarizes the results of \textit{Study A}. For SYSUSE (Fig.~\ref{fig:input_evaluation} (a)), reference image + text achieved the lowest median score (1.67), indicating the best perceived usability among the five conditions. For PEU (Fig.~\ref{fig:input_evaluation} (b)), the same modality again achieved the best median score (2.00). In contrast, single-modality inputs, especially the text-only and sketch-only conditions, showed higher medians and greater variability, suggesting increased cognitive or operational burden for non-professional users. This preference is consistent with the technical evaluation reported in Section~\ref{eval_vlm}, where reference image + text also produced the most reliable structured interpretation and downstream generation results. We therefore adopt reference image + text as the default interaction setting in the final EasyFashion system.

Fig.~\ref{fig:input_evaluation} (c) reports the results of \textit{Study B} under the selected interaction setting. Participants reported favorable experiences across all five evaluated dimensions, where lower values indicate more positive ratings: INFOQUAL (mean = 1.84, SD = 0.67), PEU (mean = 1.96, SD = 0.54), OBAU (mean = 1.93, SD = 0.75), SYSUSE (mean = 1.65, SD = 0.39), and Overall Satisfaction (mean = 1.73, SD = 0.64). These results suggest that participants found the final interaction setting both usable and satisfying in iterative design tasks, which is consistent with prior HCI work on creativity support tools arguing that such systems should be evaluated not only by output quality, but also by how well they support users' creative process and interaction experience \citep{Cherry2014}.

Qualitative feedback from the interview transcripts further supports these findings. Participants described the interaction as an iterative loop of inspecting try-on outputs, revising design intent, and regenerating results, and they reported being able to reach a satisfactory garment within a small number of refinement rounds. They particularly valued the expressive control enabled by combining a reference image with textual instructions, for example when requesting localized modifications. These observations support \textbf{H5} and \textbf{H6}, indicating that users perceived EasyFashion as easy to use and were generally able to converge on a satisfactory result efficiently.

Participants also identified several limitations. Some reported that the system tended to generate lightweight garments more successfully than layered garments, and others noted that the fabric appearance was not always sufficiently realistic, which they often attributed to the simulation choices made to maintain responsiveness. Several participants additionally requested more avatar customization options, such as selectable skin tones. Overall, these findings support the use of reference image + text as the default interaction setting for EasyFashion, while also identifying perceived authenticity as an important direction for future improvement.

Post-task interviews in \textit{Study B} further suggested that participants valued outputs that could support real-world follow-through, rather than functioning only as on-screen visualizations. For example, P6 reported that the customization process felt convenient and easy to follow, and that receiving explicit production-oriented outputs increased confidence in proceeding with personalized garment design. Several participants, such as P14, described the experience as engaging and expressed interest in using the system to customize garments aligned with their personal preferences. We also received feedback from a sewing hobbyist participant (P23), who noted that they had previously focused mainly on smaller items such as bags and accessories, but that EasyFashion's sewing pattern outputs could compensate for their lack of pattern-drafting skills, making them more confident in attempting complete garments and potentially even accepting small customization requests in the future. These findings provide support for \textbf{H7}, namely that EasyFashion can help users progress from design ideas to production-oriented garment outputs.

\begin{figure}[pos=h]
  \centering
  \includegraphics[width=\linewidth,height=.78\textheight,keepaspectratio]{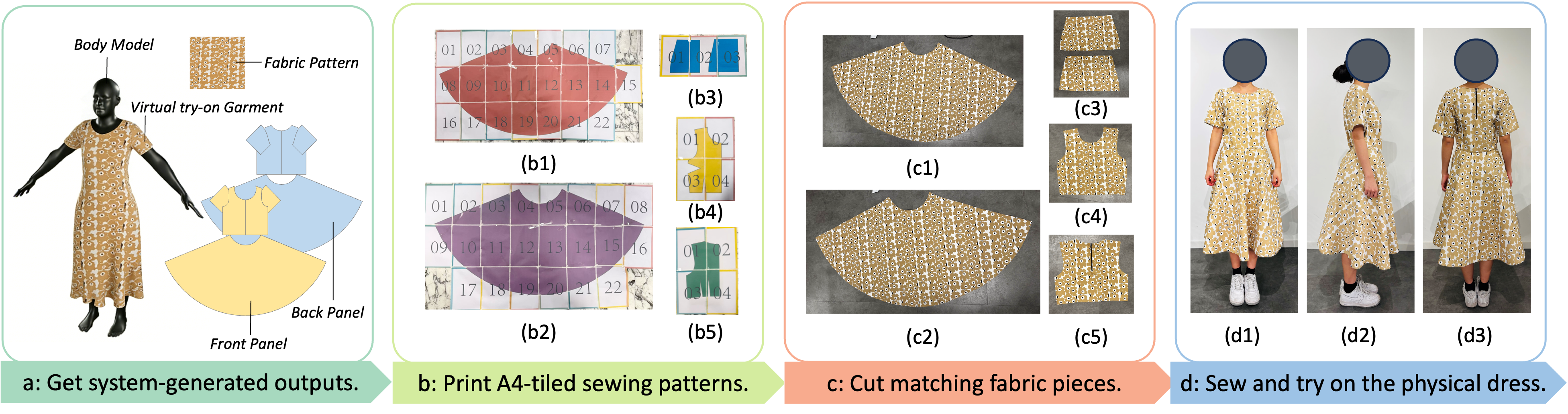}
  \caption{Illustration of a garment produced from our system outputs, from digital generation to fabric cutting, sewing, and physical try-on. (a) Get system-generated outputs, including the text-to-image fabric design, the rendered virtual try-on, and the corresponding sewing patterns. (b) Print A4-tiled sewing patterns: (b1) skirt front, (b2) skirt back, (b3) sleeve, (b4) bodice front, and (b5) bodice back. (c) Cut matching fabric pieces: (c1) skirt front, (c2) skirt back, (c3) sleeves (two pieces), (c4) bodice front, and (c5) bodice back pieces for a center zipper. Pattern pieces (b3--b5) and the corresponding fabric pieces (c3--c5) are symmetric. Therefore, we only need to place and trace one half of each pattern piece on the fabric and cut it; the other half is obtained by flipping the pattern piece and tracing/cutting again. (d) Sew and try on the physical dress: the cut fabric pieces were sewn together to form the final dress. (d1) front, (d2) side, and (d3) back, demonstrating a well-fitted result.}
  \label{fig:user_case}
\end{figure}

\subsection{Production Case Study}
\label{sec:demo_case}

Beyond the controlled evaluations reported earlier, we present a production case study to illustrate how EasyFashion can be used in practice and what the generated outputs look like in an end-to-end workflow. This example also demonstrates that the human-AI co-creation process may conclude either after several refinement rounds or after a single round when the initial output already aligns sufficiently with the user's intent.

We demonstrate this end-to-end workflow through a case in which a participant designed and produced a personalized dress using EasyFashion-generated outputs (Fig.~\ref{fig:user_case}).
In this case, the participant obtained a result that matched her intent after a single generation round and proceeded directly to production without further iterative edits. She provided input images from which the system estimated her body measurements (height 160 cm, chest 76.2 cm, waist 61 cm, hip 94 cm). These estimates were close to her self-reported measurements, providing a sanity check that the end-to-end pipeline produced plausible personalized sizing from the provided images. Based on her design intent and estimated measurements, the system generated sewing patterns for the skirt (front and back), bodice (front and back), and sleeves (Fig.~\ref{fig:user_case} (a)).

To support production, the system outputs included: (1) a text-to-image fabric design and a rendered virtual try-on for reference (Fig.~\ref{fig:user_case} (a)), (2) A4-tiled sewing patterns that she printed at true scale (Fig.~\ref{fig:user_case} (b1) to (b5)), and (3) a measurement sheet summarizing key dimensions. An independent studio tailor, who was not affiliated with the research team, produced the dress by cutting the corresponding fabric pieces (Fig.~\ref{fig:user_case} (c1) to (c5)) and assembling them, including a center zipper on the back bodice. According to the participant's report, she did not provide additional verbal explanations beyond sharing these outputs, and the tailor was able to proceed directly from the patterns and accompanying materials.

After receiving the finished dress, she reported that the overall fit was good and that the length matched her preference (Fig.~\ref{fig:user_case} (d1) to (d3)).
She also noted that off-the-shelf dresses and skirts often differ from expectations when worn, and that standard sizing can lead to awkward proportions, particularly in length, for her body. She appreciated that this workflow allowed her to choose the style and fabric herself rather than adapting to what was available on the market. At the same time, she noted that, due to limited experience, she selected a relatively stiff fabric that was not ideal for this style. Nevertheless, she still considered the process an interesting attempt and did not regret the choice.

\section{Discussion}

\subsection{Measurement Reliability for Fit Decisions}
The results partially support \textbf{H1}. The estimated measurements are most useful when fit decisions depend on broad body proportions rather than detailed local shaping. Height and hip measurements are relatively stable and visually constrained, so small estimation errors rarely affect overall sizing decisions. By contrast, chest and waist measurements are more sensitive to posture, clothing, and camera angle, and small deviations may translate into noticeable differences for close-fitting tops or waist shaping. Image-based anthropometry is known to be sensitive to viewpoint variation and landmarking error, and circumference measurements in particular may be less reliable because they are inferred rather than directly measured~\citep{MEUNIER2000445}.

This points to a practical boundary for decision support. The system can be treated as dependable for overall sizing decisions, whereas waist-sensitive details are better treated as provisional~\citep{Harvey2024}. One practical strategy is to pair the system estimate with a quick user verification of a key circumference, and then rely on iteration to correct local fit issues. This also suggests a simple interface improvement: the system could indicate which dimensions are likely to be stable and which are likely to be uncertain, so that users can better judge when to trust the estimate and when to verify it.

\subsection{Multimodal Constraints for Expressive Control}
The results support \textbf{H2}. The reference image + text condition performs well because the two inputs constrain different aspects of design intent. The image anchors silhouette, proportions, and stylistic cues that are difficult to specify precisely in words~\citep{Song2025}. The text, in turn, can clarify what the image leaves ambiguous, such as what should be preserved, what should be changed, and which constraints matter most. Together, these inputs reduce the need for the system to guess~\citep{Dritsas2025Multimodal}.

At the same time, this pairing can fail when users are unsure how to distribute intent across modalities. If the image suggests one direction and the text suggests another, the system may produce an interpretation that is reasonable but still feels unexpected to the user~\citep{ZHANG2024}. This suggests that multimodality benefits from light structure. The interface should encourage complementary roles, with the image serving as a stable reference and the text serving as a concise description of intended changes, rather than treating the two channels as interchangeable.

\subsection{Robustness for Body Diversity}
The results support \textbf{H3}. Across our demonstrations, the pipeline produces coherent outputs for different body shapes and garment types. This matters because users are more likely to continue exploring when the system behaves consistently as conditions change. Here, consistency is not only about visual quality; it also concerns whether similar edits lead to similar kinds of changes even when the underlying body shape differs~\citep{Afroogh2024TrustInAI}.

However, robustness in curated examples does not guarantee robustness in everyday use~\citep{Tocchetti2025}. User photos may vary in lighting, pose, and clothing, and garments may include edge cases such as unusual seam placements or extreme silhouettes. This makes robustness not only a technical issue but also a communication issue~\citep{Li2023}. When inputs fall outside a likely reliable range, the workflow should encourage verification or offer a conservative path that prioritizes stability over aggressive edits~\citep{10.1145/3637318}.

\subsection{Personalization Benefits for Better Fit}
The results support \textbf{H4}. The comparisons clarify why personalization can reduce the compromises associated with standard sizing. Standard size labels compress many body differences into a small number of discrete categories, so users often have to trade off length against width or tolerate localized tightness in order to achieve a desired overall silhouette. Conditioning sewing patterns on an individual's measurements reframes the task from selecting the closest standard size to fitting the specific body, which better supports decisions based on overall body proportions~\citep{wolff2023}.

At the same time, the evidence should be interpreted as suggestive rather than definitive. Pressure maps and virtual try-on results can highlight potential tight areas, but they do not fully capture comfort during movement or individual preferences for ease allowance~\citep{KimChae2025ClothingPressure}. A more cautious interpretation is that personalization improves fit plausibility and reduces obvious mismatches, while final comfort and wearability still benefit from iterative refinement and, ideally, broader wear testing beyond any single proxy measure~\citep{GarmentDiffusion2025}.

\subsection{Ease of Use for Idea Expression}
The results support \textbf{H5}. In \textit{Study A}, the reference image + text condition received the best ratings on both usability and perceived ease of use, whereas single-modality inputs showed worse median scores and greater variability. This suggests that users find it easier to express design ideas when they can begin with a concrete reference and then add a short constraint, rather than relying on a single modality to convey the full intent. \textit{Study B} shows a compatible pattern. Ratings for perceived ease of use remained favorable, and interview feedback described the interaction as a loop of inspecting try-on outputs, revising intent, and regenerating results.

Ease of expression becomes less reliable when users cannot anticipate what a given edit will change or when the controllable parameters remain unclear~\citep{Tankelevitch2024Metacognitive, 10.1145/3613904.3642466}. In such cases, users may hesitate, make excessive edits, or stop early because the output no longer provides clear guidance for the next step. This suggests that refinement support should focus on clarity during iteration. Keeping the action space small, providing examples of effective short edits, and making changes across iterations visible would help users connect revisions to outcomes without increasing upfront effort~\citep{Masson2024DirectGPT_arXiv, OTOOLE2024100080}.

\subsection{Rapid Convergence for Iterative Design}
The results support \textbf{H6}. Most participants reached an acceptable garment design within a small number of refinement rounds, suggesting that iteration functions as a practical decision-making process rather than as open-ended trial and error. Each round produces a concrete artifact for evaluation, which helps users move from vague intent to targeted adjustments~\citep{10.1145/3706598.3714316, 10.1145/3706598.3713375}. This is particularly important for non-professional users, who may struggle to specify all constraints upfront but can still reliably judge when a result is getting close to what they want~\citep{Tankelevitch2024Metacognitive}.

However, rapid convergence may reflect either efficient progress or premature stopping. In a study setting, participants may stop once the task feels complete even if the design still has unresolved issues. In real-world use, a plausible preview may also be over-trusted, leading users to accept small problems that later become costly~\citep{10.1145/3696449}. The design challenge is therefore to keep iteration lightweight while still supporting calibrated decisions. Users should be able to continue refining when necessary and verify critical fit or construction details before committing.

\subsection{Pattern Outputs for Production Readiness}
The results support \textbf{H7}. The production case study suggests that the generated sewing patterns can function as production inputs rather than remaining at the level of concept visuals~\citep{10.24963/ijcai.2025/163}. This is important because many generative systems stop at rendered visual outputs, leaving an execution gap in which users can see what they want but cannot reliably reproduce it. By providing sewing patterns and measurement summaries, EasyFashion turns an interactive design session into artifacts that a maker can follow~\citep{wolff2023}.

Production readiness, however, should not be treated as a binary property. A single successful fabrication demonstrates feasibility, but broader validity depends on how different makers interpret the outputs, how tolerant the patterns are to fabric choice and sewing skill, and how often small errors require manual correction. A careful takeaway is that the system can reduce translation effort and make remote handoff more realistic, while stronger validation should involve more garments, more materials, and more diverse production contexts.

\subsection{Design Implications for Human-AI Co-Creation}

\label{sec:design_implications}

Grounded in the Fashion Agent evaluation and user studies, we derive three HCI-oriented implications that may generalize to other human-AI co-creation systems. Across these implications, we emphasize co-creation as an iterative coordination process among user intent, an editable intermediate representation, and decision-relevant feedback.

\paragraph{\textbf{Implication 1:} Provide actionable and transferable artifacts to bridge intent and execution.}
Human-AI co-creation tools should not stop at on-screen renderings~\citep{10.24963/ijcai.2025/163}.
For non-professional users, interaction often breaks down when they need to move from a generated result to the next real-world step, such as sharing the design with an expert, placing an order, or reproducing the garment. Providing artifacts that remain understandable and usable outside the system, such as structured specifications, explicit measurements, or production-oriented templates, helps users communicate intent with less ambiguity and increases confidence that the generated result is feasible.

\paragraph{\textbf{Implication 2:} Design multimodal input as complementary constraints rather than redundant channels.}
Our results suggest that the value of multimodality lies in combining inputs that constrain different aspects of intent ~\citep{Song2025}.
In our context, reference images ground visual style and structure, while text clarifies explicit requirements and resolves ambiguity. More broadly, pairing a concrete example with concise constraints can reduce user effort, improve intent clarity, and produce more stable outcomes than relying on a single modality alone.

\paragraph{\textbf{Implication 3:} Support user agency by making iterative decisions easier and lower-risk.}
Non-professional users frequently face trade-offs that cannot be resolved by choosing among predefined options. Human-AI co-creation should therefore emphasize interactions that help users articulate goals, adjust constraints, and converge through iteration, rather than treating generation as a one-shot process. In our studies, we observed indications that participants often relied on inspecting intermediate outputs to guide subsequent edits. Rather than fully deciding and specifying everything upfront, they generated an initial result, interpreted it as feedback, and then adjusted or added constraints in later rounds ~\citep{10.1145/3696449}. This suggests a tendency toward a see-then-decide interaction loop, in which outputs function as diagnostic artifacts for next-step decisions. To support this outcome-driven loop, systems should provide decision-relevant evidence, such as fit cues or constraint-satisfaction cues, together with predictable pathways for refinement. Such support can reduce trial and error, strengthen users' sense of control and trust, and shift intent articulation from upfront specification to progressive constraint negotiation, in which users begin with coarse goals and gradually tighten constraints as feedback reveals feasibility and fit. By making decision uncertainty more legible, such feedback may also reduce perceived risk and design anxiety.

\section{Limitations and Future Work}

Despite the promising performance of our system, several limitations remain. Below, we summarize them from both user-centered interaction and technical perspectives, and outline directions for future work.

\paragraph{Interaction and user-centered limitations.}
First, users may over-rely on visual outputs such as try-on renderings when making decisions. Although such visualizations improve accessibility for non-professional users, they may also create a false sense of certainty, especially when subtle fit issues or construction constraints are not fully captured. Future work should therefore explore interaction designs that communicate uncertainty and limitations more explicitly, for example by surfacing confidence cues, constraint violations, and situations in which the system is likely to be unreliable. In addition, future systems could support comparison views across iterations so that users can better assess small changes during co-creation.

Second, non-professional users may misinterpret AI-generated results as definitive recommendations rather than editable proposals. This may lead to misunderstandings about which aspects are expected to remain stable across iterations, such as measurements derived from the body model, versus which aspects remain approximate, such as style details inferred from references or artifacts introduced by rendering and simulation. Future iterations of the system could incorporate lightweight user-facing summaries of the current garment specification, including key dimensions, garment structure, and major design assumptions, so that users can validate intent and detect mismatches early. Such summaries would also strengthen human-AI co-creation by clarifying what the system believes it is generating and what the user is approving when they choose to proceed.

Third, while multimodal inputs can improve intent clarity, they may also increase cognitive load if users are unsure what information to provide in each modality or how the modalities are combined. Future work could therefore investigate scaffolding strategies such as guided input templates, examples of effective prompts, and progressive disclosure, in which additional information is requested only when the system detects ambiguity or conflicting constraints.

\paragraph{Technical and dataset limitations.}

The accuracy of human body reconstruction can be further improved. At present, our 3D body models are generated from a small set of measurements, such as height, chest circumference, waist circumference, and hip circumference. Although these parameters capture coarse body shape, they do not account for finer-grained differences such as leg length, arm length, or thigh circumference, all of which may affect garment fit. Incorporating additional measurements, or adopting richer body-shape representations, may enable more precise reconstruction and improve fit quality. Future work should also examine robustness under common capture conditions, including variations in lighting, clothing thickness, and pose, since these factors influence whether body-specific feedback remains credible in everyday use.

In addition, the garment styles supported by our system are constrained by the coverage of the GarmentCode dataset. While EasyFashion can generate diverse examples for different body types (Fig.~\ref{fig:more_examples}), it remains limited in producing design features that are underrepresented or absent in the dataset, such as certain collar variations or advanced pattern details. Addressing this limitation will require expanding and enriching garment datasets, for example by collecting broader style categories and more fine-grained pattern annotations, in order to improve generalization and increase the range of manufacturable designs the system can support. A complementary direction is to incorporate constraint-aware editing of the structured garment specification so that users can explore more styles while remaining within manufacturable boundaries.

\section{Conclusion}

We present EasyFashion, an interactive system that enables non-professional users to design personalized garments from multimodal inputs and obtain both visual previews and production-oriented outputs. EasyFashion integrates multimodal intent understanding, personalized body reconstruction, and parametric sewing pattern generation, thereby bridging creative exploration and production-oriented workflows. Through technical evaluations, user studies, and an end-to-end production case study, we show that the system supports users in expressing and refining design intent through iteration, achieving improved fit beyond standard sizing, and supporting remote workflows through production-oriented deliverables that can be shared directly with makers. Our findings highlight the value of multimodal inputs used as complementary constraints, as well as the importance of editable intermediate representations in human-AI co-creation systems that aim to connect body-specific evaluation with manufacturable outcomes.

\section*{Acknowledgements}

This work was supported in part by the Guangdong Province Philosophy and Social Science Planning Project (No. GD24YYS04) and the Start-up Fund for New Recruits of The Hong Kong Polytechnic University (No. P0049603).

\section*{Declaration of generative AI and AI-assisted technologies}

During the preparation of this work, the authors used ChatGPT/Codex to assist with language polishing, formatting checks, LaTeX troubleshooting, and preparation of submission materials. The authors also used generative AI-assisted tools to create the explanatory illustrations in Fig.~\ref{fig:storying} and Fig.~\ref{fig:placeholder}. In the research workflow, Gemini-2.5-Flash was used as the vision-language model for the Fashion Agent, as described in the Methods section. After using these tools, the authors reviewed and edited the content as needed and take full responsibility for the content of the publication.

\end{document}